\documentclass[fleqn,usenatbib]{mnras}
\usepackage{newtxtext,newtxmath}

\usepackage[T1]{fontenc}
\usepackage{ae,aecompl}

\usepackage{graphicx}	
\title[BRITE observations of $\nu$~Cen and $\gamma$~Lup]{BRITE observations of $\bnu$ Centauri and $\bgamma$ Lupi, the first 
non-eclipsing members of the new class of nascent binaries\thanks{Based on data collected by the BRITE Constellation satellite 
mission, designed, built, launched, operated and supported by the Austrian Research Promotion Agency (FFG), the University of 
Vienna, the Technical University of Graz, the University of Innsbruck, the Canadian Space Agency (CSA), the University of Toronto 
Institute for Aerospace Studies (UTIAS), the Foundation for Polish Science \& Technology (FNiTP MNiSW), and National Science 
Centre (NCN).}}

\author[M.~Jerzykiewicz et al.]
{M.~Jerzykiewicz,$^{1}$\thanks{E-mail: mjerz@astro.uni.wroc.pl} 
A.~Pigulski,$^{1}$ 
G.~Michalska,$^{1}$ 
D.~Mo\'zdzierski,$^{1}$
M.~Ratajczak,$^{1,2}$\newauthor
G.~Handler,$^{3}$
A.F.J.~Moffat,$^{4}$
H.~Pablo,$^{5}$
A.~Popowicz,$^{6}$
G.A.~Wade$^{7}$
 and
K.~Zwintz$^{8}$
 \\  
$^{1}$Astronomical Institute, University of Wroc{\l}aw, Kopernika 11, 51-622 Wroc{\l}aw, Poland\\ 
$^{2}$Astronomical Observatory, University of Warsaw, Al. Ujazdowskie 4, 00-478 Warsaw, Poland\\
$^{3}$Nicolaus Copernicus Astronomical Center, Polish Academy of Sciences, Bartycka 18, 00-716 Warsaw, Poland\\
$^{4}$D\'ept. de physique, Univ. de Montréal, C.P. 6128, Succ. C-V., Montr\'eal, QC, H3C 3J7 
and Centre de Recherche en Astrophysique du Qu\'ebec, Canada\\
$^{5}$American Association of Variable Star Observers, 49 Bay State Road, Cambridge, MA 02138, USA\\
$^{6}$Silesian University of Technology, Department of Electronics, Electrical Engineering and Microelectronics, Akademicka 16, 
44-100 Gliwice, Poland\\
$^{7}$Department of Physics and Space Science, Royal Military College of Canada, PO Box 17000, Station Forces, Kingston, ON, 
Canada, K7K 0C6\\
$^{8}$Institut f\"ur Astro- und Teilchenphysik, Universit\"at Innsbruck, Technikerstra{\ss}e 25, A-6020 Innsbruck
} 

\date{Accepted XXX. Received YYY; in original form ZZZ}

\pubyear{2021}

\begin{document}
\label{firstpage}
\pagerange{\pageref{firstpage}--\pageref{lastpage}}
\maketitle

\begin{abstract} 
Results of an analysis of the BRITE-Constellation and SMEI photometry and radial-velocity observations, archival and new, of 
two single-lined spectroscopic binary systems $\nu$~Centauri and $\gamma$~Lupi are reported. In the case of $\gamma$ Lup AB, a 
visual binary, an examination of the light-time effect shows that component A is the spectroscopic binary. Both $\nu$~Cen and 
$\gamma$~Lup exhibit light variations with the orbital period. The variations are caused by the reflection effect, i.e.~heating of 
the secondary's hemisphere by the early-B main sequence (MS) primary component's light. The modelling of the light curves 
augmented with the fundamental parameters of the primary components obtained from the literature photometric data and {\em 
Hipparcos\/} parallaxes, shows that the secondary components are pre-MS stars, in the process of contracting onto the MS. 
$\nu$ Cen and $\gamma$ Lup A are thus found to be non-eclipsing counterparts of the B2\,IV eclipsing binary (and a $\beta$ Cephei 
variable) 16 (EN) Lac, the B5\,IV eclipsing binary (and an SPB variable) $\mu$ Eri, and the recently discovered LMC nascent 
eclipsing binaries.
\end{abstract}


\section{Introduction}\label{intro}
Formation of close binaries with massive components is a challenging theoretical problem as the details of the formation processes 
cannot be tested observationally. This is because massive stars at the pre-MS stage of evolution are embedded in the parent cloud 
from which the stellar cluster is formed \citep[e.g.][]{2003ARA&A..41...57L}. The stars become observable only after the cloud is 
dispersed by strong stellar winds of the most massive stars or core-collapse supernova explosions. Assuming coevality or 
near-coevality of the onset of star formation, one comes to the conclusion that if the binary has a low mass ratio $q=M_2/M_1$, 
the system may become visible as a binary consisting of a massive MS primary and a pre-MS companion. Given the overall scarcity 
of massive stars, their fast evolution and observational selection effects, such systems will be rare and difficult to discover. 
Recently, using a sample of 174\,000 eclipsing binaries in the Large Magellanic Cloud (LMC), selected from the third phase of the 
Optical Gravitational Lensing Experiment (OGLE-III) database \citep{Gra+11}, \cite{MoDiS15} discovered 18 systems with low 
mass-ratios ($0.06 < q < 0.25$), consisting of early B-type MS primaries with $M_1\approx 5$\,--\,16\,M$_{\sun}$ and pre-MS 
secondaries. After correcting the observed number of these systems for selection effects, they found that (2.0\,$\pm$\,0.6)\% of 
early B-type MS stars would have companions with masses of 0.06--0.25\,$M_1$ and orbital periods, $P_{\rm orb}$, of 3.0--8.5~d. 
According to \cite{MoDiS15}, this fraction is $\approx$10 times larger than that observed around solar-type MS stars in the same 
mass ratio and period interval.

Such systems, referred to by \cite{MoDiS15} as nascent eclipsing binaries with extreme mass ratios (NEBs henceforth), are 
interesting because it is not known how they were formed. In particular, there is the question whether these systems originate as 
a close binary from the beginning and both components accrete matter at the same time \citep[see e.g.][and references 
therein]{2018arXiv180806488S} or if they form as a wide binary and the orbit shrinks at a later stage of evolution leaving a much 
closer system as a product \citep*[e.g.][]{1998MNRAS.300..292K,2007ApJ...669.1298F}. Equally interesting is the future of these 
systems after the primary leaves the MS: the stars may coalesce or form a low-mass X-ray binary 
\citep{1998ApJ...493..351K,2006MNRAS.369.1152K}. Further evolution may produce a type Ia supernova and a millisecond pulsar.

As discussed by \citet{2017ApJS..230...15M}, early-type binaries identified by various observing techniques, such as spectroscopy, 
eclipses, interferometry, adaptive optics, common proper motion, etc., are distributed in distinct regions of the $q$--$P_{\rm 
orb}$ plane. Those with low-mass pre-MS secondaries are difficult to select because of the large difference in brightness between 
components. In the case of a wide binary observed by means of astrometric techniques, the pre-MS status of the secondary is hard 
to establish due to the limited possibilities of deriving its mass and radius. The systems with the shortest orbital periods can 
be detected as spectroscopic binaries (SBs) showing a measurable reflection effect. If there are no eclipses, an analysis of the 
reflection light-curves may constrain the radius of the secondary component, revealing its pre-MS status. In the present paper, we 
announce the discovery of two such systems, $\nu$~Cen and $\gamma$~Lup, non-eclipsing counterparts of the NEBs. The presence of 
eclipses would be an additional asset allowing to constrain the secondary's radius even if the reflection effect is marginal or 
below the detection threshold. Examples illustrating this are the single-lined (SB1) eclipsing binaries 16 (EN) Lac and $\mu$ Eri. 
The B2\,IV primary of the former is also a well-known $\beta$ Cephei variable, while the B5\,IV primary of the latter is an SPB 
star. The pre-MS nature of the secondary of 16 (EN) Lac was suggested by \cite{PJ88} and confirmed by \cite{MJ+15}, while that of 
$\mu$ Eri is established in Section\,\ref{discussion} of the present paper.

\section{Target stars: $\bnu$~Cen and $\bgamma$~Lup}\label{targets}
$\nu$~Cen (HD\,120307, HR\,5190, HIP\,67464) is a member of the Upper Centaurus Lupus (UCL) subgroup of the association Sco~OB2 
\citep{1999AJ....117..354D}. The MK type is B2\,IV \citep*{HGS69}. According to \citet*{W15}, $\nu$~Cen is an SB1 system with 
$P_{\rm orb}=2.625172$\,d, a circular orbit and the semi-amplitude $K=20.6$\,km\,s$^{-1}$. These parameters were confirmed by 
\cite*{ASR85} but \citet{Raj77} obtained $e=0.26$ and $K =29.9$\,km\,s$^{-1}$. In addition, \citet{Raj77} maintains that the 
residuals from the orbital radial-velocity (RV) curve show a $\beta$~Cephei-type variation with at least two periods close to 
0.1750\,d. This period, albeit a single one, was found by \citet{KS82} in their RV and Str\"omgren $u$-filter observations. Using 
residuals from the orbital solution and the RV data of \citet*{KS82}, \cite{ASR85} updated this period to 0.1690156\,d. 
Subsequently, \citet{AP92} added three nights of RV observations and revised the period to 0.1696401\,d. However, \citet{Sh78}, 
\citet*{PJM81} and \citet{SJ83} failed to detect any short-period brightness variations in their Str\"omgren $b$-filter 
observations. Moreover, from a number of 12\,{\AA}\,mm$^{-1}$ spectrograms, \citet{SJ83} found no evidence for a short-period RV 
variation. More recently, from a series of high-resolution CCD spectrograms, \citet{ST02} discovered a pattern of moving bumps in 
the Si\,III 455.2 and 456.7-nm line profiles which they attributed to high-degree non-radial pulsations with $l$ ranging 
from 6 to 10, ruling out low-degree pulsations with a period of 0.17\,d. A small-amplitude brightness variation of $\nu$~Cen with 
the orbital period was detected by \citet{WR83}; these authors ascribed the variation to a reflection effect. From numerous 
observations in the Str\"omgren $b$ filter, \citet*{Cuy+89} concluded that
\begin{quote} 
The observations [...] clearly show a variation with the same period as the orbital period. Minimum light corresponds to maximum 
positive RV. As discussed by \citet{WR83}, the resulting light curve is most probably a reflection effect. A careful investigation 
did not reveal any signs of a variation at any other frequency with an amplitude exceeding 2~millimags.
\end{quote} 

In the General Catalogue of Variable Stars (GCVS)\footnote{http://www.sai.msu.su/gcvs/gcvs/}, $\nu$~Cen is classified as BCEP 
(i.e.~a $\beta$~Cephei-type variable) but \citet{StaHan05} have degraded it to the status of a candidate $\beta$~Cephei variable 
presumably because of the conflicting evidence for short-period RV and brightness variability summarized above. In the {\em 
Hipparcos\/} catalogue \citep{ESA97}, the type of variability is EB (i.e.~a $\beta$ Lyrae-type eclipsing variable), the range is 
3.318\,--\,3.329\,mag, and the period $P=2.6249\,\pm$\,0.0003\,d.

$\gamma$~Lup (HD\,138690, HR\,5776, HIP\,76297), is a close visual double with the separation ranging from 0.1 to 0.8 arcsec. It 
consists of components of very nearly equal brightness: the Hp magnitudes are equal to 3.397 and 3.511 for the A and B 
component, respectively \citep{ESA97}. The orbital elements listed in the US Naval Observatory's Sixth Catalog of Orbits of Visual 
Binary Stars (ORB6)\footnote{http://www.usno.navy.mil/USNO/astrometry/optical-IR-prod/wds/orb6}, computed by \citet{Hz90} from the 
1836\,--\,1988 mainly micrometric observations, include a period of 190\,yr, a semi-major axis of 0.655\,arcsec, an 
inclination of 95\fdg0 and an eccentricity of 0.51. In Notes to the 5th edition of the Bright Star Catalogue \citep{HW91}, an MK 
type of B2\,IV-V is assigned to either component, but \citet{HGS69} give a single MK type of B2\,IV. One component is an SB1 
system with $P_{\rm orb} =2.80805$\,d and $e =0.10\,\pm\,$0.02 \citep{Lev+87}. These authors note that both components fell on the 
slit of the spectrograph, so that---because of their similar brightness---either can be responsible for the RV variation. The star 
is listed in the GCVS as ELL: (i.e.~a questionable ellipsoidal variable). According to the {\em Hipparcos\/} catalogue, the type 
of variability is P (i.e.~periodic), the range is 2.693\,--\,2.711\,mag, and the period, $P ={}$2.8511$\,\pm\,$0.0004\,d. Note 
that the frequencies corresponding to the {\em Hipparcos\/} and the spectroscopic period differ by 0.0054\,d$^{-1} 
={}$2\,yr$^{-1}$. An excellent summary of the observations of $\gamma$~Lup throughout 1987 was provided by \citet{Baa87}.

\section{The data and reductions}
\subsection{BRITE and SMEI photometry}\label{sect:brite}
\begin{figure} 
\centering
\includegraphics[width=\columnwidth]{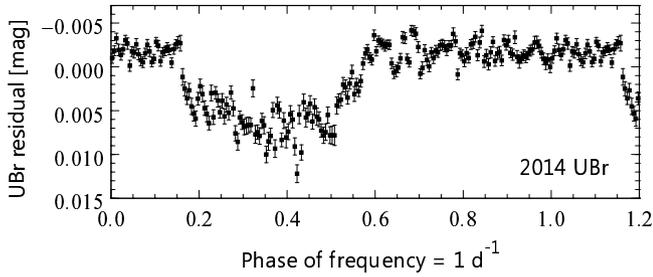} 
\caption{The spurious 1\,d$^{-1}$ variation of the 2014 UBr magnitudes of $\gamma$~Lup. Shown are the normal points computed in 
adjacent intervals of 0.005 phase of the frequency of 1\,d$^{-1}$. The data were the 2014 UBr magnitudes of $\gamma$~Lup 
decorrelated with respect to the CCD's temperature, the $X$ and $Y$ coordinates of the image centre of gravity and the satellite 
orbital phase, and pre-whitened with the star's orbital frequency.}
\label{Fig-01} 
\end{figure}

The photometry analysed in the present paper was obtained from space by the constellation of five BRITE (BRIght Target 
Explorer) nanosatellites \citep{BRITE1,2016PASP..128l5001P} during two runs, in the fields Centaurus I (both stars) and Scorpius I 
(only $\gamma$~Lup). These observations were secured by all five BRITEs, three red-filter, Uni\-BRITE (UBr), BRITE-Toronto (BTr), 
and BRITE-Heweliusz (BHr), and two blue-filter ones, BRITE-Austria (BAb) and BRITE-Lem (BLb). Details of BRITE observations are 
given in Table \ref{Tab-01}. The Cen\,I observations were obtained in stare mode, the Sco I, in the chopping mode of observing 
\citep{2016PASP..128l5001P,2017A&A...605A..26P}. The images were analyzed by means of two pipelines described by 
\cite{2017A&A...605A..26P}. The resulting aperture photometry is subject to several instrumental effects \citep{2018adlc106} and 
needs pre-processing aimed at the effects' removal. To remove the instrumental effects we followed the procedure designed by 
\cite{AP+16} with several modifications proposed by \cite{2018adlc175}. The whole procedure includes converting fluxes to 
magnitudes, rejecting outliers and the worst orbits (i.e.~the orbits on which the standard deviation of the magnitudes was 
excessive), and one- and two-dimensional decorrelations with all parameters provided with the data (e.g.~position in the subraster 
and CCD temperature) and the calculated satellite orbital phase. The 2014 UBr observations of $\gamma$~Lup were additionally 
decorrelated with respect to the frequency of 1\,d$^{-1}$ because they showed a spurious variation with this frequency (see 
Fig.\,\ref{Fig-01}). After decorrelating, the magnitudes were de-trended and deviant magnitudes were rejected by hand.

\begin{table*}
\caption{Details of BRITE and SMEI data for $\nu$~Cen and $\gamma$~Lup. $N_{\rm orig}$ and $N_{\rm final}$ are the original and 
final (after pre-processing) numbers of data points. RSD is the residual standard deviation after subtracting the stars' intrinsic 
variability.}
\label{Tab-01}
\begin{tabular}{cccccrrrr}
\hline
Star & Field & Satellite &  Start & End & \multicolumn{1}{c}{Length of} & \multicolumn{1}{c}{$N_{\rm orig}$} & 
\multicolumn{1}{c}{$N_{\rm final}$} & \multicolumn{1}{c}{RSD} \\
       &   &                  &     date   &  date   & \multicolumn{1}{c}{the run [d]} &&&\multicolumn{1}{c}{[mmag]}\\
\hline
$\nu$~Cen & Cen I & BAb & 2014.04.09 & 2014.08.18 & 131.4 & 39\,863 & 37\,256  & 11.98\\
 & & BLb & 2014.06.12 & 2014.07.08 & 26.6 & 3\,979 & 3\,929  & 6.78\\
 & & UBr & 2014.03.25 & 2014.08.17 & 145.3 & 65\,621 & 60\,461  & 12.17\\
 & & BTr & 2014.06.27 & 2014.07.03 & 6.0 & 4\,949 & 4\,909  & 5.54\\
$\gamma$~Lup & & BAb & 2014.04.09 & 2014.08.18 & 131.3 & 39\,814 & 36\,841  & 11.29\\
 & & BLb & 2014.06.12 & 2014.07.08 & 26.6 & 4\,011 & 3\,491  & 14.33\\
 & & UBr & 2014.03.25 & 2014.08.17 & 145.3 & 64\,255 & 60\,698  & 12.07\\
 & & BTr & 2014.06.27 & 2014.07.03 & 6.0 & 4\,963 & 4\,918  & 5.90\\
\hline
$\gamma$~Lup &  Sco I & BAb & 2015.03.28 & 2015.07.19 & 112.9 &4\,972 &  3\,269 & 7.32\\
& & BLb & 2015.03.19 & 2015.08.26 & 160.6 & 40\,152 & 35\,695 & 5.74\\
&  & UBr & 2015.03.20 & 2015.08.29 & 162.1 & 58\,341 & 54\,135 & 10.74\\
&  & BHr & 2015.06.26 & 2015.08.28 & 63.8 & 20\,904 & 7\,563 & 8.59\\
\hline
$\nu$~Cen& & SMEI & 2003.02.02 & 2010.12.30 & 2888.0 & 28\,187 & 19\,643  & 7.06\\
$\gamma$~Lup & & SMEI & 2003.02.02 & 2010.11.16 & 2843.2 & 27\,193 & 22\,034 & 11.11\\
 \hline
\end{tabular}
\end{table*}

The {\em Solar Mass Ejection Imager\/} ({\sc SMEI}) time-series photometry has been already employed by a number of workers. Eg., 
\citet{Kal+17} used {\sc SMEI} time-series photometry to bolster the BRITE data analysis of V389~Cygni, a triple system with an 
SPB star component. The time-series photometry of $\nu$~Cen and $\gamma$~Lup was downloaded from the SMEI 
website\footnote{http://smei.ucsd.edu/new$_-$smei/data\&images/stars/time\-series.html} and reduced as described in section 6 of 
\citet{Kal+17}. The details of the photometry are given in Table \ref{Tab-01}.

\subsection{The radial-velocity data}\label{RVs}
For both target stars, several sets of archival RV data are available in the literature. They are listed in Table \ref{Tab-02} as 
sets (1) to (15). In addition to sets (1) to (10) of the archival RVs of $\nu$~Cen, and sets (11) to (15) of $\gamma$~Lup, we used 
three archival FEROS spectrograms of $\gamma$~Lup, obtained in May 2004 and August 2013, to derive the RVs. This was done by 
least-squares fitting of rotationally-broadened template spectra to the observed spectra. The template spectra were calculated 
using BSTAR2006 grid of models \citep{2007ApJS..169...83L}, convolved rotationally with the {\tt rotin3} program provided on the 
{\tt Synspec}\footnote{http://nova.astro.umd.edu/Synspec43/synspec.html} web page. These RVs, together with their estimated 
standard errors, are listed in Table \ref{Tab-03} and are referred to in Table \ref{Tab-02} as set (16). In addition to the 
archival data, new RVs of $\gamma$~Lup were derived from spectrograms taken by one of the authors (MR) with the BACHES 
echelle spectrograph \citep{koz2016} installed on the 0.5-m robotic telescope Solaris-1 \citep{koz2014,syb14} located at SAAO. 
Eight spectrograms each of exposure time of 250~s and signal-to-noise ratios from 100 to 150 were taken between 29 February and 5 May 
2016. In order to reduce the spectrograms, we used standard IRAF\footnote{IRAF is distributed by the National Optical Astronomy 
Observatory, which is operated by the Association of Universities for Research in Astronomy (AURA) under a cooperative agreement 
with the National Science Foundation.} procedures. The wavelength calibration was performed using the mean of wavelength solutions 
from the Th-Ar lamp frames taken before and after target frames. IRAF \texttt{rvsao.bcvcorr} task was used for barycentric 
velocity and time corrections. The RVs were calculated as straight means of the values obtained from Gaussian fitting to four 
He\,I lines, 438.79, 492.19, 587.56, and 667.82\,nm. Standard errors of these means serve as a measure of uncertainty. The 
new RVs together with the uncertainties are listed in Table~\ref{Tab-03} and are referred to in Table \ref{Tab-02} as set (17). 

\begin{table} 
\caption{A log of RV measurements of $\nu$~Cen and $\gamma$~Lup.} 
\label{Tab-02} 
\centering  
\begin{tabular}{ccrl} 
\hline 
Set & Year(s) & RVs & Source\\ 
\hline 
\multicolumn{4}{c}{$\nu$~Centauri}\\ 
\hline 
(1) & 1904\,--\,1907 & 11 & \cite*{W15} \\ 
(2) & 1914 & 9 &  \cite*{W15} \\ 
(3) & 1968\,--\,1973 & 37 & \cite{Raj77}, H and He lines\\ 
(4) & 1974\,--\,1976 & 12 & \cite{Lev+87}\\ 
(5) & 1979 & 49 & \cite{SJ83} \\ 
(6) & 1980 & 26 & \cite{KS82}\\ 
(7) & 1983\,--\,1984 & 21 & \cite{ASR85}\\ 
(8) & 1985\,--\,1988 & 53 & \cite{AP92}\\ 
(9) & 1998 & 93 & \cite{ST02}, Si\,III lines\\ 
(10) & 2002 & 3 & \cite{Ji+06}\\ 
\hline 
\multicolumn{4}{c}{$\gamma$~Lupi}\\ \hline (11) & 1914\,--\,1917 & 10 & \cite{CM28}\\ (12) & 1954\,--\,1957 & 2 & \cite{BM60}\\ 
(13) & 1955 & 6 & \citet*{vH+63}\\ 
(14) & 1966 & 19 & \cite{vAS69}\\ 
(15) & 1974\,--\,1976 & 8 & \citet{Lev+87}\\ 
(16) & 2004\,--\,2013 & 3 & this paper, FEROS; see Table \ref{Tab-03}\\ 
(17) & 2016 & 8 & this paper, BACHES; see Table \ref{Tab-03}\\ 
\hline 
\end{tabular} 
\end{table}

\begin{table}
\caption{The FEROS (F) and BACHES (B) RVs of $\gamma$~Lup.}
\label{Tab-03}
\begin{center}
\begin{tabular}{crcr}
\hline
\multicolumn{1}{c}{HJD$-$}&\multicolumn{1}{c}{RV} & \multicolumn{1}{c}{HJD$-$}&\multicolumn{1}{c}{RV}\\
\multicolumn{1}{c}{2450000} &\multicolumn{1}{c}{[km\,s$^{-1}$]} & \multicolumn{1}{c}{2450000} &\multicolumn{1}{c}{[km\,s$^{-1}$]}\\
\hline
3126.9094&0.3 $\pm$ 2.1 (F) & 7470.4093&$-$7.2 $\pm$ 3.5 (B)\\
6525.5973&23.2 $\pm$ 1.5 (F) & 7497.4058&25.3 $\pm$ 7.0 (B)\\
6525.5984&25.8 $\pm$ 1.8 (F) & 7498.3859&21.2 $\pm$ 6.9 (B)\\
7448.4996&9.8 $\pm$ 2.3 (B) & 7513.4071&$-$13.5 $\pm$ 4.2 (B)\\
7450.4967&$-$8.9 $\pm$ 0.8 (B) & 7514.3792&37.9 $\pm$ 7.6 (B)\\
7469.4180&41.8 $\pm$ 4.3 (B) &&\\
\hline
\end{tabular}
\end{center}
\end{table}

\section{The orbital RV and light curves of $\bnu$ Cen}
\subsection{The orbital period}\label{porb-ncen}
We shall now derive the orbital period of $\nu$~Cen from the available RV and photometric observations. The RV data sets we could 
use, i.e.~those in which the measurements were well distributed over orbital phase, include six sets listed in Table \ref{Tab-02}: 
(1), (2), (3), (4), (7), and (9). For set (9) straight means of the RVs of the Si\,III 455.2 and 456.7\,nm lines 
\citep{ST02} were taken. We fitted each data set with the sine curve
\begin{equation}\label{Eq-01}
{\rm RV} = \gamma + K_1 \sin (2 \pi f_{\rm orb} t + \phi_i), 
\end{equation} 
where $f_{\rm orb} ={}$0.38092\,d$^{-1}$ and $t$ is reckoned from the middle of the interval spanned by the observations.  
Using a sine curve is justified because the orbit of $\nu$ Cen is circular (see Section 4.2). In fitting Eq.~(\ref{Eq-01}), we 
used the method of least squares with weights inversely proportional to the squares of the standard errors of the velocities. In 
sets (1) and (2), the standard errors were obtained by multiplying the overall probable error given by \citet*{W15} by 
$\sqrt{\langle n \rangle/n}/$0.6745, where $n$ is the number of lines measured; if two values of $n$ were listed by \citet*{W15}, 
a mean was taken, if none (one case), $n$ was set equal to $\langle n \rangle$, the overall mean of $n$. In sets (3) and (9), the 
standard errors were assumed to be those of an equal-weight fit of Eq.~(\ref{Eq-01}) to the data. In sets (4) and (7), the 
standard errors were computed from the probable errors given by \citet{Lev+87} and \cite{ASR85}, respectively. Then, we computed 
the epochs of crossing the $\gamma$-axis from the smaller to greater RV (i.e.~from approach to recession), HJD$_{\gamma}$, and 
their standard deviations. These numbers are given in the second column of Table~\ref{Tab-04}.

The photometric observations of $\nu$~Cen we used consisted of the following sets: 1987 and 1988 $b$ magnitudes of \citet{Cuy+89}, 
{\em Hipparcos\/} Hp magnitudes, SMEI data divided into five adjacent segments of approximately equal duration, and BRITE blue and 
red magnitudes (see Section\,\ref{sect:brite}). We fitted the magnitudes $m$ with the sine curve
\begin{equation}\label{Eq-02}
m = \langle m\rangle + A \sin (2 \pi f_{\rm orb} t + \phi), 
\end{equation}
with $f_{\rm orb} ={}$0.38092\,d$^{-1}$ and $t$ reckoned from the middle of the interval spanned by the observations; in the case 
of the Hp and BRITE fits, we applied weights inversely proportional to the squares of the standard errors of the magnitudes. The 
epochs of maximum light computed from the fits, HJD$_{\rm max}$, and their uncertainties are listed in the first column of 
Table~\ref{Tab-05}. Assuming that at a given epoch the orbital phase of HJD$_{\gamma}$ is equal to that of HJD$_{\rm max}$, an 
assumption to be verified shortly, and using HJD$_{\gamma}$ from Table~\ref{Tab-04} and HJD$_{\rm max}$ from Table~\ref{Tab-05} we 
arrived at the following ephemeris:
\begin{equation}\label{Eq-03}
{\rm HJD}_{\gamma} = {\rm HJD}\,2450893.6665(25) + 2.6252541(22) \times E.
\end{equation} 
The number of cycles that elapsed from the $E=0$ epoch and the residuals O\,--\,C from the ephemeris are listed in 
Tables~\ref{Tab-04} and \ref{Tab-05}. In order to verify our assumption that the orbital phase of HJD$_{\gamma}$ is equal to that 
of HJD$_{\rm max}$, we computed the $E=0$ epoch using the RV and brightness data separately. The results, 
HJD\,2450893.6678$\,\pm\,$0.0015 and 2450893.696$\,\pm\,$0.013, differ by about 2$\sigma$. We conclude that at maximum light the 
secondary component is at superior conjunction, i.e. the primary component is the closest to the observer, with the secondary 
behind. This phase relation between the RV and light variation is compatible with the reflection effect. Note that this conclusion 
is in conflict with that of \citet{Cuy+89} quoted in Section~\ref{targets}.

\begin{table}
\centering
\caption{The epochs of crossing the $\gamma$-axis from the smaller to greater RV, HJD$_{\gamma}$, the number of cycles, $E$, and 
the residuals O\,--\,C from ephemeris (3).}
\label{Tab-04}
\begin{tabular}{ccrrc}
\hline
Set & \multicolumn{1}{c}{HJD$_{\gamma}-$2400000}&\multicolumn{1}{c}{$E$}&\multicolumn{1}{r}{O\,--\,C [d]}\\
\hline
(1) &  17180.0780\,$\pm$\,0.0206 & $-$12842 & $-$0.0754\\
(2) &  20325.0512\,$\pm$\,0.0487 & $-$11644 & $-$0.1566\\
(3) & 40851.9605\,$\pm$\,0.0667 & $-$ 3825 & $-$0.1091\\
(4) & 42545.4276\,$\pm$\,0.0528 & $-$ 3180 & $+$0.0691\\
(7) & 45590.6037\,$\pm$\,0.0879 & $-$ 2020 & $-$0.0495\\
(9) & 50893.6678\,$\pm$\,0.0011 &        0 &  $+$0.0013\\
\hline
\end{tabular}
\end{table}

\begin{table}
\centering
\caption{The epochs of maximum light, HJD$_{\rm max}$, the number of cycles, $E$, and the residuals O\,--\,C from ephemeris (3).}
\label{Tab-05}
\begin{tabular}{@{}crrc@{}}
\hline
\multicolumn{1}{c}{HJD$_{\rm max}-$2400000}&\multicolumn{1}{c}{$E$}&\multicolumn{1}{r}{O\,--\,C [d]}&Data\\
\hline
46971.8720\,$\pm$\,0.0768 & $-$1494 &    0.3351&\citet{Cuy+89} \\
47286.5900\,$\pm$\,0.0312 & $-$1374 &    0.0226&\citet{Cuy+89} \\
48501.9848\,$\pm$\,0.0410 &  $-$911 & $-$0.0752&{\em Hipparcos\/}\\
52962.3911\,$\pm$\,0.0080 &     788 &    0.0244&SMEI\\
53539.9307\,$\pm$\,0.0079 &    1008 &    0.0081&SMEI\\       
54135.8685\,$\pm$\,0.0073 &    1235 &    0.0132&SMEI\\       
54697.6682\,$\pm$\,0.0076 &    1449 &    0.0085&SMEI\\       
55272.5796\,$\pm$\,0.0084 &    1668 & $-$0.0107&SMEI\\       
56821.4760\,$\pm$\,0.0053 &    2258 & $-$0.0143&BRITE blue\\       
56821.4822\,$\pm$\,0.0027 &    2258 & $-$0.0081&BRITE red\\
\hline
\end{tabular}
\end{table}

\begin{figure*} 
\includegraphics[width=0.8\textwidth]{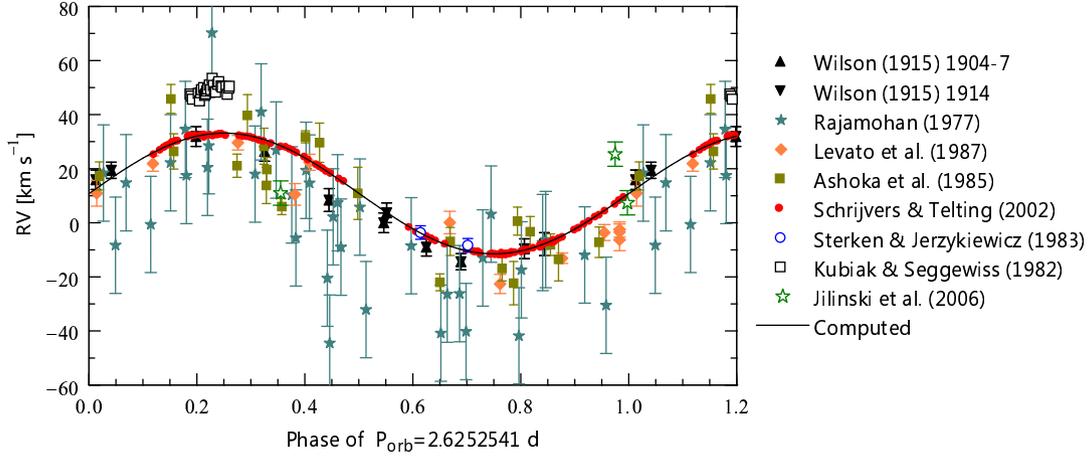} 
\caption{Archival RVs of $\nu$~Cen plotted as a function of orbital phase. The phases were computed from ephemeris (3). The solid 
line is the RV curve computed from the spectroscopic elements listed in Table~\ref{Tab-06}.}
\label{Fig-02} 
\end{figure*}

\subsection{The RV curve and the spectroscopic orbit}
The RVs of $\nu$~Cen are plotted in Fig.\,\ref{Fig-02}. For the \citet*{ST02} points, the error bars (not shown) would be of about 
the same size as the symbols plotted in the figure because we assumed the standard error to be equal to 0.6\,km\,s$^{-1}$, the 
standard deviation of the least-squares fit of the sine-curve to these points (see Section\,\ref{porb-ncen}). In addition to the 
data referenced in Table~\ref{Tab-04} (filled symbols), the RV measurements of \citet{SJ83}, \citet{KS82} and \citet{Ji+06}, i.e. 
sets (5), (6) and (10) detailed in Table \ref{Tab-02}, are shown (open symbols). In the case of \citet{SJ83}, we plotted only the 
first and the last of their 49 data-points with the standard errors provided by them. In the case of \citet{Ji+06}, we converted 
the MJD of the exposure beginning (which they call HJD in their table 1) to HJD of the middle of the exposure by adding 
2400000.5\,d, the heliocentric correction and half of the exposure time. In addition, we estimated the standard error from the 
range of the RVs on HJD\,2452413 (the two asterisks at phase $\sim$0.99 in Fig.\,\ref{Fig-02}) to be 4.5\,km\,s$^{-1}$. As can be 
seen from the figure, most measurements scatter around those of \citet{ST02}; the two exceptions are the measurements of 
\citet*{Raj77} and \citet*{KS82}. The former show a systematic shift toward smaller RVs, the latter, toward greater RVs. Omitting 
the \citet*{Raj77} and \citet*{KS82} data, we computed a spectroscopic orbit by means of the non-linear least squares method of 
\citet{s} with weights inversely proportional to the squares of the standard errors of the velocities. The parameters of the orbit 
are determined by RVs of \citet{ST02} but their standard deviations, by the remaining data. The eccentricity of the orbit turned 
out to be an insignificant $e=0.009\,\pm\,0.007$. We conclude that the orbit is circular, confirming the result of \citet*{W15}. 
The elements of the orbit are listed in Table~\ref{Tab-06} and the RV curve computed from these elements is shown in 
Fig.\,\ref{Fig-02} with the solid line.

\begin{table}
\centering
\caption{Orbital elements of $\nu$~Cen.}
\label{Tab-06}
\begin{tabular}{cc}
\hline
Orbital period, $P_{\rm orb}$&2.6252541 d (assumed)\\
Epoch of crossing the gamma axis\\ from approach to recession, HJD$_{\gamma}$&2450893.6658\,$\pm$\,0.0031\\
Eccentricity, $e$&\ 0 (assumed)\\
$\gamma$ velocity&\ 10.77\,$\pm$\,0.09 km\,s$^{-1}$ \\
Semi-amplitude of primary's RVs, $K_1$&\ 22.30\,$\pm$\,0.12 km\,s$^{-1}$ \\
Projected semi-major axis, $a_1\sin i$&\ 1.157\,$\pm$\,0.006 R$_{\sun}$\\
Mass function, $f(M)$&\ 0.00302\,$\pm$\,0.00005 M$_{\sun}$\\
\hline
\end{tabular}
\end{table}

\subsection{The orbital light-curves and the W-D modelling}\label{wd-ncen}
\begin{figure} 
\includegraphics[width=\columnwidth]{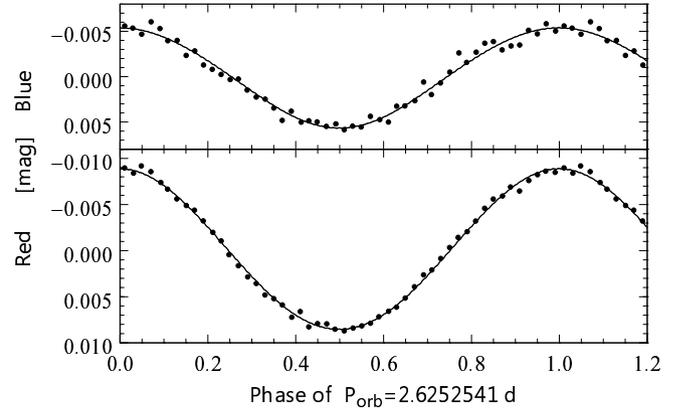} 
\caption{The light-curves of $\nu$~Cen. Plotted are normal points formed in the adjacent intervals of 0.02 orbital phase from the 
blue (upper panel), and red (lower panel) BRITE magnitudes. The phases were computed from ephemeris (\ref{Eq-03}). The lines are 
the theoretical W-D light-curves.}
\label{Fig-03} 
\end{figure}

The light-curves of $\nu$~Cen are presented in Fig.\,\ref{Fig-03}. The data shown are normal points formed in adjacent intervals 
of 0.02 orbital phase from the blue and red BRITE magnitudes (the upper and lower panel, respectively). The error bars are not 
shown because they would barely extend beyond the plotted circles: the standard errors ranged from 0.27 to 0.42\,mmag for the blue 
normal points, and from 0.26 to 0.39\,mmag for the red normal points. The lines plotted in the figure are the theoretical 
light-curves, results of the Wilson-Devinney (W-D) modelling to be discussed presently. A least-squares fit of a sum of the 
$f_{\rm orb}$ and 2$f_{\rm orb}$ sines to the normal points yields blue amplitudes of 5.47$\,\pm\,$0.22 and 
0.11$\,\pm\,$0.22\,mmag, and red amplitudes of 8.70$\,\pm\,$0.16 and 0.14$\,\pm\,$0.16\,mmag, respectively. The 2$f_{\rm orb}$ 
amplitudes are consistent with the conclusion of Appendix \ref{ap-nc} that no detectable ellipsoidal light-variation is present.
Thus, the orbital light-variation is caused solely by the reflection effect. 

The light curves were subject to modelling by means of the 2015 version of the W-D 
code\footnote{ftp://ftp.astro.ufl.edu/pub/wilson/lcdc2015/}  \citep{1971ApJ...166..605W,1979ApJ...234.1054W}. In addition to 
the light curves, the input data for the code included the orbital elements from Table~\ref{Tab-06} and the primary component's 
fundamental parameters $T_{\rm eff,1} ={}$22\,370~K and $\log g_1 ={}$3.76 from Appendix \ref{param-nc}. The $T_{\rm eff,1}$ was 
fixed throughout but the final $\log g_1$, used only in deriving the limb darkening coefficients, were taken from the W-D 
solutions after several iterations. The limb darkening coefficients were from the logarithmic-law tables of Walter V.~Van 
Hamme\footnote{http://faculty.fiu.edu/{$\sim$}vanhamme/limb-darkening/, see also \citet{VH}.}. For the primary component we 
assumed [M/H] $= 0$ and used 421 and 620.5\,nm monochromatic coefficients for the blue and red data, respectively. A 
radiative-envelope albedo of 1.0 was assumed. In treating the reflection effect, we used the detailed model with six reflections 
(MREF $=2$, NREF $=6$). For the secondary component, bolometric limb-darkening coefficients and a convective-envelope albedo of 
0.5 were adopted.

The mass function, $f(M)$, depends on the primary's mass, $M_1$, the secondary's mass, $M_2$, and the inclination, $i$, or --- 
alternatively --- on $M_1$, $i$, and the mass ratio $q = M_2/M_1$. Once two of these parameters are fixed, the third can be 
calculated from $f(M)$ given in Table \ref{Tab-06}. The evolutionary mass $M_1=8.7\,\pm\,$0.3\,M$_{\sun}$ derived for 
the primary component of $\nu$~Cen in Appendix \ref{param-nc} is model dependent. Therefore, in order to be certain that the 
range of $M_1$ adopted in the W-D modelling comprises the true primary mass, we assumed a range five times wider than the formal 
uncertainty of $M_1$, i.e.~7.2 to 10.2\,M$_{\sun}$. In a preliminary run we found that the range of $i$ should be limited to 
$30\degr\leq i <75\degr$ because for $i \geq 75\degr$ the W-D light-curves showed an eclipse not seen in the observed 
light-curves, while for $i<30\degr$ an ellipsoidal effect, also not present in the observed light-curves, showed up in the W-D 
light curves. In the $i<30\degr$ models, the distortion of the primary causing the ellipsoidal effect was a consequence of a 
higher $M_2$ and a tighter orbit. In addition, a higher $M_2$ and hence greater brightness would be also inconsistent with 
the lack of the secondary's lines in the spectrum. The final W-D modelling was therefore done assuming a grid of ($M_1$, $i$) 
$\in$ (7.2\,--\,10.2~M$_{\sun}$, 35$\degr$\,--\,75$\degr$). As it turned out, the $q$ calculated for the whole grid was confined 
to a rather narrow range of 0.074\,--\,0.169.

\begin{figure} 
\includegraphics[width=\columnwidth]{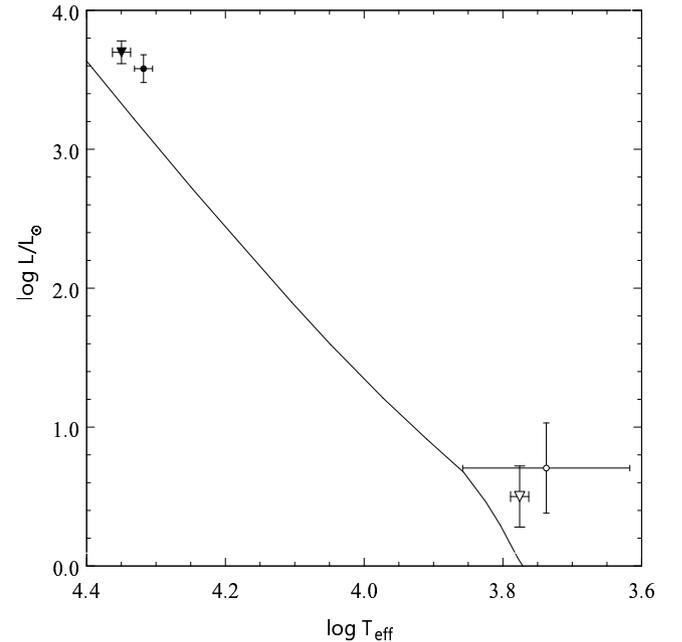} 
\caption{The components of $\nu$ Cen and $\gamma$ Lup A (inverted triangles and circles, respectively) in the HR diagram. The line 
is the ZAMS from \citet{2012A&A...537A.146E} for the $Z=0.014$ models with no rotation. For $\log(L_1/\mbox{L}_{\sun}) < 1.0$, the 
ZAMS for models with rotation is virtually identical with the one shown.}
\label{FigHRpreMS}
\end{figure}

In the ($M_1$, $i$) grid, we fitted the W-D light-curves to the normal points shown in Fig.~\ref{Fig-03} assuming equal weights. 
The overall standard deviation of the fits were very nearly the same over the whole grid. This is a consequence of the fact that 
there are three free parameters in the models, the secondary's effective temperature, $T_{\rm eff,2}$, and the components' radii, 
$R_1$ and $R_2$ (formally, the surface potentials), that can be adjusted to get a satisfactory fit. Thus, the fits do not 
discriminate between different ($M_1$, $i$). The solid lines plotted in Fig.\,\ref{Fig-03} are the W-D light-curves computed with 
$M_1 ={}$8.695~M$_{\sun}$ and $i=35\degr$. At the resolution of the figure, the light-curves computed with other ($M_1$, $i$) 
values would be impossible to distinguish from those shown. The ranges of the parameters of the components obtained from the W-D 
modelling are given in Table~\ref{Tab-07}; they are an order of magnitude greater than the formal standard deviations of the 
W-D solutions. It is interesting that the primary's W-D surface gravity is much better constrained than that derived in Appendix 
\ref{param-nc} from the photometric indices. In Fig.~\ref{FigHRpreMS}, the components of $\nu$ Cen are plotted using the effective 
temperature and luminosity of the primary from Appendix \ref{param-nc}, and those of the secondary from Table~\ref{Tab-07}; the 
ranges of $\log T_{\rm eff,2}$ and $\log(L_2/\mbox{L}_{\sun})$ listed in the table define the full lengths of the error bars and 
the open inverted triangles are placed at their intersection. Also shown in the figure is the ZAMS for the $Z=0.014$ models from 
\cite{2012A&A...537A.146E}. Given coevality of the components, the position of the secondary relative to the ZAMS indicates its 
pre-MS status. The secondary's evolutionary status will be further discussed in Section\,\ref{discussion}.

\begin{table}
\centering
\caption{The ranges of parameters of the components of $\nu$~Cen and $\gamma$~Lup~A obtained from the W-D modelling for the given 
ranges of the mass of the primary.}
\label{Tab-07}
\begin{tabular}{ccc}
\hline
Parameter & $\nu$~Cen & $\gamma$~Lup~A\\
\hline
$M_1$ [M$_{\sun}$]  (assumed) & $\langle$7.2\,--\,10.2$\rangle$& $\langle$6.0\,--\,10.0$\rangle$\\
\hline
$M_2$ [M$_{\sun}$] &$\langle$0.59, 1.45$\rangle$&$\langle$0.72, 1.93$\rangle$\\
$R_1$ [R$_{\sun}$] & $\langle$3.93, 4.56$\rangle$& $\langle$3.92, 5.39$\rangle$\\
$R_2$ [R$_{\sun}$] & $\langle$1.30, 2.10$\rangle$&$\langle$2.00, 3.47$\rangle$\\
$T_{\rm eff,2}$ [K] & $\langle$5790, 6150$\rangle$& $\langle$4140, 7210$\rangle$\\
$M_{\rm bol,1}$ & $\langle$$-$4.43, $-$4.10$\rangle$&$\langle$$-$4.47, $-$3.78$\rangle$\\
$M_{\rm bol,2}$ &$\langle$$+$2.94, $+$4.05$\rangle$&$\langle$$+$2.17, $+$3.78$\rangle$\\
$\log(L_1/\mbox{L}_{\sun})$ &$\langle$3.54, 3.67$\rangle$ &$\langle$3.41, 3.68$\rangle$\\
$\log(L_2/\mbox{L}_{\sun})$ & $\langle$0.28, 0.72$\rangle$&$\langle$0.38, 1.03$\rangle$\\
$\log(g_1/(\mbox{cm\,s}^{-2}))$ &$\langle$4.103, 4.132$\rangle$ &$\langle$3.867, 4.130$\rangle$\\
$\log(g_2/(\mbox{cm\,s}^{-2}))$ &$\langle$3.965, 3.987$\rangle$ &$\langle$3.515, 3.803$\rangle$\\
$L_3/(L_1+L_2)_{\rm blue}$ & 0.0 (assumed) & $\langle$0.31, 1.46$\rangle$\\
$L_3/(L_1+L_2)_{\rm red}$ & 0.0 (assumed) & $\langle$0.34, 2.18$\rangle$\\
\hline
$q=M_2/M_1$ & $\langle$0.074, 0.169$\rangle$& $\langle$0.099, 0.234$\rangle$\\
$a$ [R$_{\sun}$] & $\langle$16.0, 18.2$\rangle$&$\langle$16.0, 19.3$\rangle$ \\
\hline
\end{tabular}
\end{table}

\section{The triple system of $\bgamma$ Lup}
\subsection{The SB1 RV curve}\label{sb1rv}
All sets of the RV observations of $\gamma$~Lup detailed in Section\,\ref{RVs} and Table \ref{Tab-02} can be phased with the 
photometric orbital frequency of 0.35090\,d$^{-1}$ provided that one day is added to all JD epochs of observations of 
\citet{vH+63} (their UT dates are correct) and a misprint in one epoch of observation of \citet{vAS69} is corrected: in their 
table 3, April 25.019 should be replaced with April 25.919. In four sets, (11), (14), (15), and (17), the RV measurements are 
distributed in orbital phase sufficiently well to be fit with the sine curve in which $f_{\rm orb}=0.35090$\,d$^{-1}$. Using 
a sine curve is justified because the SB1 orbit of $\gamma$ Lup is circular (see Section 5.4). In fitting Eq.\,(\ref{Eq-01}), we 
used the method of least squares with weights inversely proportional to the squares of the standard errors of the velocities. For 
sets (11) and (14), the standard errors were obtained by multiplying the standard error of an equal-weights fit by $\sqrt{\langle 
n \rangle/n}$, where $n$ is the number of lines measured and $\langle n \rangle$ is the overall mean of $n$. For set (15), the 
standard errors were computed from the probable errors provided by \citet{Lev+87}. For set (17), the standard errors are given in 
Table~\ref{Tab-03}. The fits yielded the $\gamma$ velocities to be used in Section\,\ref{casb}, and the epochs of crossing the 
$\gamma$-axis from the smaller to greater RV (i.e.~from approach to recession), HJD$_{\gamma}$, to be used in Section\,\ref{lite}.

\subsection{Component A is the SB}\label{casb} 
\begin{figure*}
\includegraphics[width=0.8\textwidth]{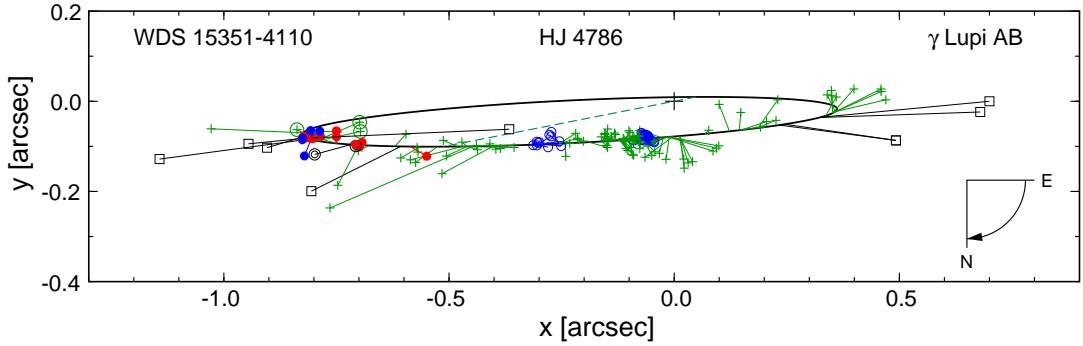} 
\caption{The apparent relative orbit of the $\gamma$~Lup AB (HJ\,4786) system. The ellipse corresponds to the solution given in 
Table~\ref{Tab-08}. The individual measurements are plotted with different symbols, depending on the observing technique. The 
techniques were the following: micrometric measurements (green pluses; the three discovery measurements made by J.\ F.\ W.\ 
Herschel in 1835\,--\,1837 are encircled), visual interferometry (open blue circles), speckle interferometry (filled red circles), 
CCD, mostly lucky imaging (filled blue circles), {\it Hipparcos\/} and Tycho measurements (black double circles). The data 
excluded from the fit are plotted as black squares. All measurements are connected with the calculated position in the orbit, 
corresponding to the epoch of observation. The large plus marks the position of component A and the dashed line is the line of 
apsides.}
\label{Fig-04} 
\end{figure*}

Given the elements of the $\gamma$~Lup AB visual binary orbit, the parallax of the system and the mass ratio of the 
components, the temporal variation of the RVs of the components A and B can be computed. After \citet{Hz90} derived the elements 
mentioned in Section\,\ref{targets}, a number of interferometric determinations of the position angle $\theta$ and angular 
separation $\rho$ of $\gamma$~Lup B relative to A became available. In order to update the orbital elements, we compiled a list of 
$\theta$ and $\rho$ from the US Naval Observatory's Fourth Catalog of Interferometric Measurements of Binary Stars\footnote{ 
http://www.usno.navy.mil/USNO/astrometry/optical-IR-prod/wds/int4}, assigning weights to the measurements depending on the number 
of observations used in computing $\theta$ and $\rho$, the scatter of the observations and the telescope size. Zero weight was 
given to the measurements with $\rho$ grossly deviating from the average run; this was never the case for $\theta$. The updated 
elements were estimated by bootstrapping with 1\,000 resamplings. The elements are listed in Table~\ref{Tab-08} and the 
corresponding apparent relative orbit is plotted in Fig~\ref{Fig-04}. The value of $a ={}0\farcs$970 from the table and the 
revised {\em Hipparcos\/} parallax, equal to 7.75$\,\pm\,$0.50 mas \citep{vL}, yield the semi-major axis of the relative orbit 
$a=125.2$\,AU. Inserting this value and the period of the AB system $P_{\rm AB}=167.3$\,yr into Kepler's third law, we get 
the mass of the system $M_{\rm A} + M_{\rm B}=70.1$\,M$_{\sun}$, a value much too large for a pair of B2 stars. The lower bound of 
the orbital period, $P_{\rm AB}=160.1$\,yr, yields a value still greater. Using the upper bound of the orbital period allowed by 
the solution, $P_{\rm AB}=188.8$\,yr, and the corresponding $a=1.394\times 10^{10}$\,km, we get $M_{\rm A} + M_{\rm 
B}=22.7$\,M$_{\sun}$, a value still 5.5~M$_{\sun}$ greater than the overall mass of the system derived in Section~\ref{param-gl}. 

\begin{figure} 
\includegraphics[width=\columnwidth]{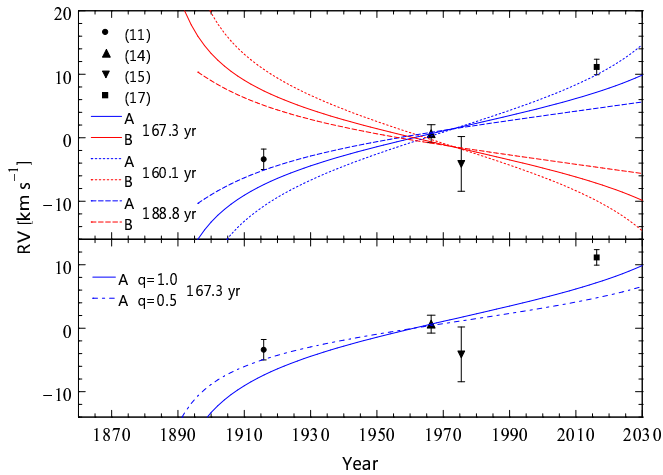} 
\caption{The RVs of the components A and B of $\gamma$~Lup, computed from the orbital elements of Table~\ref{Tab-08} and the 
revised {\em Hipparcos\/} parallax, for the B to A mass ratio $q=1.0$ (upper panel), and the RV of component A for $q=1.0$ and 0.5 
(lower panel) compared with the $\gamma$ velocities determined in Section\,\ref{sb1rv} from the four RV sets detailed in the text 
(symbols with error bars).}
\label{Fig-05} 
\end{figure}

Taking the above-mentioned three values of $P_{\rm AB}$ and the corresponding $T$, $e$, $i$ and $\omega$ from Table~\ref{Tab-08}, 
and assuming $\gamma ={}$0.00~km\,s$^{-1}$, we computed the RV curves shown in Fig.\,\ref{Fig-05}. Also plotted are the $\gamma$ 
velocities from the fits carried out in Section\,\ref{sb1rv} decreased by 4.12\,km\,s$^{-1}$, so that sets' (14) $\gamma$ 
(triangle) coincides with the computed RVs of component A for $P_{\rm AB}=167.3$\,yr and $q=1.0$ (blue solid line). It is clear 
from the figure that (i) the $\gamma$ velocities approximately follow the computed RV variation of component A, justifying this 
section's heading, and (ii) the available RV data are insufficient to constrain the elements of the visual binary orbit or the 
component's mass ratio. 

\begin{table}
\centering
\caption{Parameters of the visual binary orbit of $\gamma$~Lup AB (HJ 4786).}
\label{Tab-08}
\begin{tabular}{cc}
\hline
Parameter & Value\\
\hline
Orbital period, $P_{\rm AB}$ [yr] &167.3$^{+21.5}_{-7.2}$\\[1mm]
Time of periastron passage, $T$ [yr] &1885.7$^{+2.9}_{-4.0}$\\[1mm]
Semi-major axis, $a$ [arcsec] &0.970$^{+0.343}_{-0.240}$\\[1mm]
Eccentricity, $e$&0.826$^{+0.082}_{-0.181}$\\[1mm]
Inclination, $i$ [$\degr$]&93.04$^{+1.45}_{-0.89}$ \\[1mm]
Longitude of periastron, $\omega$ [$\degr$]&286.90$^{+12.0}_{-5.9}$ \\[1mm]
Position angle of the line of nodes, $\Omega$ [$\degr$]&91.20$^{+0.52}_{-0.35}$ \\
\hline
\end{tabular}
\end{table}

\subsection{The light-time effect}\label{lite}
Because of the orbital motion around the centre of mass of the AB system, the epochs of observations of $\gamma$~Lup A include a 
term arising from the light-time effect (LiTE). In particular, the epochs of crossing the $\gamma$-axis from the smaller to 
greater RV, HJD$_{\gamma}$, derived in Section\,\ref{sb1rv}, and the epochs of maximum light, HJD$_{\rm max}$, to be derived 
shortly, will be affected. HJD$_{\gamma}$ are listed in the second column of Table~\ref{Tab-09} while HJD$_{\rm max}$, in the 
first column of Table \ref{Tab-10}. The latter were computed from the least-squares fits of Eq.\,(\ref{Eq-02}) with $f_{\rm 
orb}=0.35090$\,d$^{-1}$ to the {\em Hipparcos\/}, SMEI, and BRITE data. The SMEI data were divided into five adjacent segments of 
approximately equal duration. Before fitting, the 2014 BRITE magnitudes were pre-whitened with the 1.5389\,d$^{-1}$ term, and the 
2015 blue magnitudes were pre-whitened with the 1.6913\,d$^{-1}$ term derived in Appendix \ref{ap-gl3}. 

\begin{table}
\caption{The epoch of crossing the $\gamma$-axis from the smaller to greater RV, HJD$_{\gamma}$, the number of cycles, $E$, and 
the residuals (O\,--\,C)$_{1.0}$ and (O\,--\,C)$_{0.5}$ from the $P_{\rm AB}=188.8$\,yr, $q=1.0$ and 0.5 ephemerides (see 
Section\,\ref{lite}).}
\label{Tab-09}
\begin{tabular}{crrcc}
\hline
Set & \multicolumn{1}{c}{HJD$_{\gamma}-$2400000}&\multicolumn{1}{c}{$E$}&\multicolumn{1}{r}{(O\,--\,C)$_{1.0}$}&
\multicolumn{1}{c}{(O\,--\,C)$_{0.5}$}\\
\hline
(11) & 20811.868\,$\pm$\,0.039 & $-$12868 &$-$0.117 & $-$0.015\\
(14) &39255.415\,$\pm$\,0.026 &  $-$6396 &$-$0.269 & $-$0.204\\
(15) & 42566.952\,$\pm$\,0.116 &  $-$5234 &$-$0.163 & $-$0.104\\
(17) & 57482.791\,$\pm$\,0.023 &        0 &$-$0.009 & $+$0.020\\
\hline
\end{tabular}
\end{table}

\begin{figure} 
\includegraphics[width=\columnwidth]{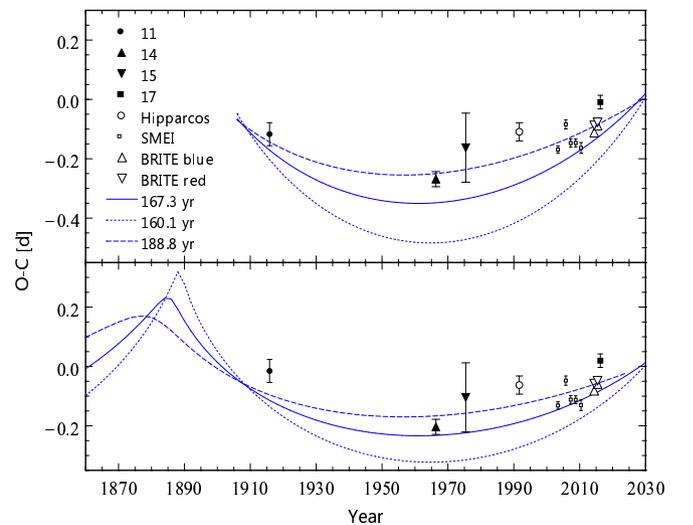} 
\caption{The LiTE O\,--\,C curves for $\gamma$~Lup A, computed from the orbital elements of Table~\ref{Tab-08} (lines) for the B 
to A mass ratio $q =1.0$ (upper panel) and 0.5 (lower panel), compared with the (O\,--\,C)$_{1.0}$ and (O\,--\,C)$_{0.5}$ 
residuals from Tables~\ref{Tab-09} and \ref{Tab-10} (symbols with error bars).}
\label{Fig-06} 
\end{figure}

In Fig.\,\ref{Fig-06} there are shown the LiTE O\,--\,C curves computed for $\gamma$~Lup A from equation (3) of \citet{Irw52} with 
the 167.3, 160.1 and 188.8-yr elements of Table~\ref{Tab-08} (the solid, short-dashed and dashed lines, respectively) for two 
values of the B to A mass ratio, $q=1.0$ (upper panel) and 0.5 (lower panel). Assuming that at a given epoch the orbital phase of 
HJD$_{\gamma}$ is equal to that of HJD$_{\rm max}$, i.e. that the maximum light of the SB light-curve occurs when the secondary is 
at superior conjunction, one can fit the LiTE O\,--\,C curves to the HJD$_{\gamma}$ and HJD$_{\rm max}$ from Tables~\ref{Tab-09} 
and \ref{Tab-10}. In the ephemerides $T_0 + P_0 E$ obtained in this way, the period $P_0$ is equal to the $\gamma$~Lup SB A 
orbital period, $P_{\rm orb}$, for the epoch of $d({\rm O}-{\rm C})/dE=0$. For the 188.8-yr elements, which yield the smallest 
value of $M_{\rm A} + M_{\rm B}$ we consider (see Section\,\ref{casb}), the residuals computed with $q=1.0$ and 0.5, 
(O\,--\,C)$_{1.0}$ and (O\,--\,C)$_{0.5}$, are listed in Tables~\ref{Tab-09} and \ref{Tab-10} and plotted in Fig.\,\ref{Fig-06}. 
The $q=1.0$ residuals fit the computed O\,--\,C curve with a slightly smaller standard deviation (0.0071\,d) than the $q=0.5$ 
residuals (0.0076\,d). 

\begin{table}
\centering
\caption{The epoch of maximum light, HJD$_{\rm max}$, the number of cycles, $E$, and the residuals (O\,--\,C)$_{1.0}$ and 
(O\,--\,C)$_{0.5}$ from the $P_{\rm AB}=188.8$~yr, $q=1.0$ and 0.5 ephemerides (see Section\,\ref{lite}).}
\label{Tab-10}
\begin{tabular}{crccc}
\hline
\multicolumn{1}{c}{HJD$_{\rm max}$}&\multicolumn{1}{c}{$E$}&\multicolumn{1}{r}{(O\,--\,C)$_{1.0}$}&\multicolumn{1}{r}
{(O\,--\,C)$_{0.5}$}&Source of\\$[{\rm HJD}-2400000]$ &&[d]&[d]&data\\
\hline
48500.2220(306) & $-$3152 & $-$0.1093 & $-$0.0626&{\em Hipparcos\/}\\
52999.9459(125) & $-$1573 & $-$0.1690 & $-$0.1314&SMEI\\
53649.7780(156) & $-$1345 & $-$0.0840 & $-$0.0477&SMEI\\
54199.7201(142) & $-$1152 & $-$0.1472 & $-$0.1119&SMEI\\
54749.7261(133) &  $-$959 & $-$0.1464 & $-$0.1122&SMEI\\
55351.0101(180) &  $-$748 & $-$0.1634 & $-$0.1305&SMEI\\
56815.8423(083) &  $-$234 & $-$0.1118 & $-$0.0818&BRITE blue\\
56815.8671(057) &  $-$234 & $-$0.0877 & $-$0.0576&BRITE red\\
57180.6374(047) &  $-$106 & $-$0.0886 & $-$0.0593&BRITE blue\\
57180.6457(051) &  $-$106 & $-$0.0774 & $-$0.0481&BRITE red\\
\hline
\end{tabular}
\end{table}

\subsection{The SB RV and the mean light-curves of $\bgamma$ Lup A}\label{sbrvmlc}
Using the $P_{\rm AB}=188.8$\,yr, $q=1.0$ LiTE O\,--\,C curve (dashed line in the upper panel of Fig.\,\ref{Fig-06}), we corrected 
the epochs of the RV and the BRITE observations of $\gamma$~Lup for the LiTE, i.e.~reduced the epochs to the centre of mass of the 
AB system. In addition, we reduced all RV measurements using component's A computed RVs (blue dashed line in the upper panel of 
Fig.\,\ref{Fig-05}) and removed a systematic difference of 6.4\,km\,s$^{-1}$ between sets (14) and (17). Then, we re-determined 
HJD$_{\gamma}$ and HJD$_{\rm max}$. The ephemeris obtained from these data
\begin{equation}\label{Eq-04}
{\rm HJD}_{\gamma} = {\rm HJD}\,2457482.7996(39) + 2.8497690(47) \times E.
\end{equation}
will be unaffected by the orbital motion of component A provided that the true $P_{\rm AB}$ and $q$ are equal to those we 
assumed. The initial epoch and the period in ephemeris (\ref{Eq-04}) are close to those of the $P_{\rm AB}=188.8$\,yr, $q=1.0$ 
ephemeris derived in Section\,\ref{lite}. This comes about because $d{\rm RV}/dt=0$ at the epoch of $d({\rm O}-{\rm C})/d{\rm 
E}=0$.

\begin{figure*} 
\includegraphics[width=0.82\textwidth]{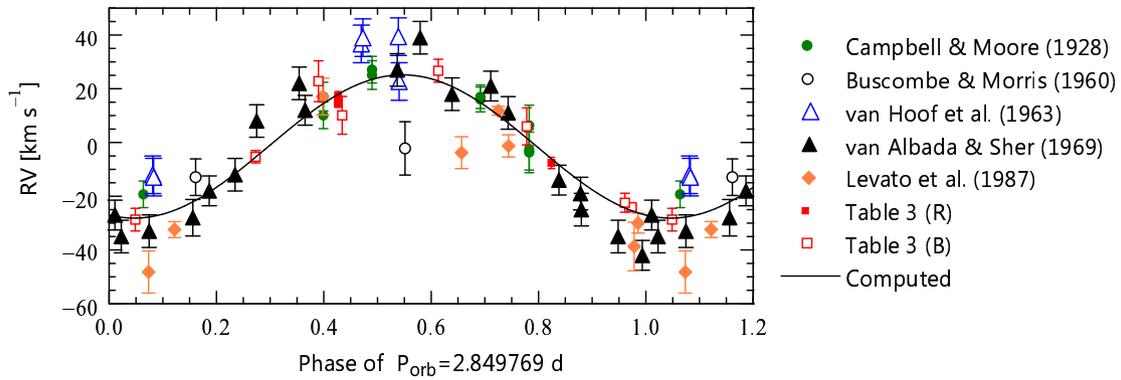} 
\caption{The RVs of $\gamma$~Lup, corrected for the LiTE as explained in the text, plotted as a function of the orbital phase. The 
epoch of phase zero is HJD\,2457482. The solid line is the RV curve computed from the spectroscopic elements listed in 
Table~\ref{Tab-11}.}
\label{Fig-07} 
\end{figure*}

The reduced RVs are plotted in Fig.\,\ref{Fig-07} as a function of phase computed using the LiTE-corrected epochs of observations 
and $P_{\rm orb}$ from ephemeris (\ref{Eq-04}). These data were then used in computing a spectroscopic orbit by means of the 
non-linear least squares method of \citet{s} with weights inversely proportional to the squares of the standard errors. The 
standard errors of sets (11), (14), (15), and (17) were the same as in Section\,\ref{sb1rv} while those of set (16) were taken 
from Table~\ref{Tab-03}. For set (12), we obtained the standard errors from the probable errors \citet{BM60} provide, while for 
set (13) we estimated the standard error from the scatter in the phase diagram. The eccentricity of the orbit turned out to be 
equal to an insignificant $e=0.044\,\pm\,$0.045. It is thus feasible to assume that the orbit is circular. The elements of a 
circular orbit are listed in Table~\ref{Tab-11} and the RV curve computed from these elements is shown in Fig.\,\ref{Fig-07} with 
the solid line. 

An anonymous referee has suggested that the orbital period of the AB binary could be obtained from an assumed value of the mass of 
the system equal to a sum of masses of two B2 stars. We carried out this exercise using the mass of the system of 
17.2$\,\pm\,$0.7~M$_{\sun}$ derived in Section~\ref{param-gl} and the same data as in the first paragraph of Section~5.2. We 
obtained $P_{\rm AB} ={}$198.3$\,\pm\,$5.4~yr and the remaining orbital elements rather close to those of \citet{Hz90} mentioned 
in Section~\ref{targets}. If plotted in the upper panel of Fig.~\ref{Fig-05}, the RVs computed with the $P_{\rm AB} ={}$198.3-yr 
elements would very nearly coincide with those computed with the $P_{\rm AB} ={}$188.8-yr elements. A similar result is obtained 
for the $O-C$ shown in Fig.~\ref{Fig-06}. In addition, the elements of the $\gamma$ Lup~A spectroscopic orbit obtained from the 
HJD and RV corrected for LiTE with the $P_{\rm AB} ={}$198.3-yr elements differ from those in Table~\ref{Tab-11} by much less than 
1$\sigma$: $K_1$, $a_1 \sin i$ and $f(M)$ differ by 0.5$\,\pm\,$1.4~km\,s$^{-1}$, 0.03$\,\pm\,$0.08~R$_{\sun}$ and 
0.0003$\,\pm\,$0.0008~M$_{\sun}$, respectively. Clearly, there exists an interval of mass of $\gamma$ LupAB such that for a mass 
from this interval there is a value of $P_{\rm AB}$ which accounts for the observed temporal variation of RV and $O-C$. We believe 
that more measurements of $\theta$ and $\rho$ of the system are needed to narrow down this interval. Whether the model-independent 
mass of the system derived from the observed orbit will then agree with the model-dependent mass obtained in 
Section~\ref{param-gl} remains to be seen.

\begin{table}
\centering
\caption{Orbital elements of $\gamma$~Lup A computed under assumption of zero eccentricity from the LiTE-corrected RV shown in 
Fig.\,\ref{Fig-07}. HJD$_{\gamma}$ is the epoch of crossing the $\gamma$-axis from the smaller to greater RVs (i.e.~from approach 
to recession).}
\label{Tab-11}
\begin{tabular}{cc}
\hline
Orbital period, $P_{\rm orb}$&2.849769 d (assumed)\\
Epoch of crossing $\gamma$ axis, HJD$_{\gamma}$&2457482.846\,$\pm$\,0.021 d\\
Eccentricity, $e$&0 (assumed)\\
$\gamma$ velocity&$-$1.5\,$\pm$\,0.7 km\,s$^{-1}$ \\
Semi-amplitude of primary's orbit, $K_1$&26.7\,$\pm$\,1.0 km\,s$^{-1}$ \\
Projected semi-major axis, $a_1\sin i$&1.50\,$\pm$\,0.05 R$_{\sun}$\\
Mass function, $f(M)$&0.0056\,$\pm$\,0.0006 M$_{\sun}$\\
\hline
\end{tabular}
\end{table}

Let us now turn to the question whether correcting the epochs of observations for the LiTE would affect results of frequency 
analysis of the BRITE data. As can be seen from Fig.\,\ref{Fig-06}, the LiTE corrections to the epochs of observations over the 
time interval covered by the BRITE data can be expressed by a linear function of time, $a+bt$, where $a$ and $b$ are constants. In 
other words, the LiTE-corrected epochs of observations are shifted by $a$ and scaled by $1+b$. From the well-known properties of 
the Fourier transform it follows that the time shift translates in the frequency domain into a phase shift, while the scaling, 
into scaling the frequencies and amplitudes by the reciprocal of $1+b$. Since $b \approx 1.5\times 10^{-5}$, the answer to our 
question is no. More precisely, the LiTE corrections would have negligible effect on the frequencies and amplitudes derived in 
Appendices \ref{ap-gl1} and \ref{ap-gl3}.

\begin{figure} 
\includegraphics[width=\columnwidth]{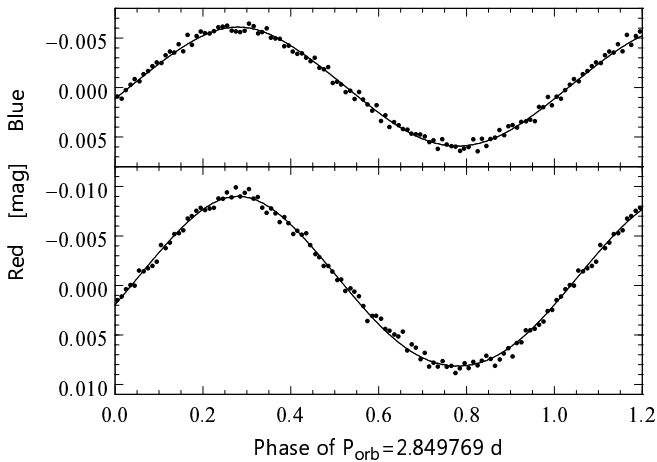} 
\caption{The orbital light-curves of $\gamma$~Lup. Plotted are normal points formed in the adjacent intervals of 0.01 orbital 
phase from the combined 2014 and 2015 blue (upper panel) and red (lower panel) BRITE magnitudes. The epoch of phase zero is 
HJD\,2457482. The lines are the theoretical W-D light-curves computed with $M_1 ={}$8.093~M$_{\sun}$ and $i=55\degr$.}
\label{Fig-08} 
\end{figure}

Using the 2014 and 2015 blue and red magnitudes (see Appendix \ref{ap-gl3}) with the LiTE-corrected epochs of observations and 
$P_{\rm orb}$ from ephemeris (\ref{Eq-04}), we plot the blue and red phase diagrams in Fig.\,\ref{Fig-08}. The data plotted in the 
figure are normal points, computed in the adjacent intervals of 0.01 orbital phase. The standard errors, ranging from 0.25 to 
0.35\,mmag for the blue normal points and from 0.24 to 0.36\,mmag for the red normal points, are not shown. The solid lines are 
the theoretical light-curves, results of the W-D modelling detailed in Section\,\ref{wd-glup}. A least-squares fit of a sum of the 
$f_{\rm orb}$ and 2$f_{\rm orb}$ sines to the normal points yields the blue amplitudes of 5.92$\,\pm\,$0.10 and 
0.14$\,\pm\,$0.10\,mmag, and the red amplitudes of 8.54$\,\pm\,$0.12 and 0.44$\,\pm\,$0.12\,mmag. In the latter case, the phase 
difference between the $f_{\rm orb}$ and 2$f_{\rm orb}$ sines is equal to 0.21$\,\pm\,$0.14\,rad, excluding ellipsoidal effect as 
the cause of the 2$f_{\rm orb}$ term. We conclude that the orbital light-variation is caused solely by the reflection effect. The 
amplitudes and the phase difference agree with the results presented in Section\,\ref{ap-gl3}. 

\subsection{The W-D modelling}\label{wd-glup}
In the W-D modelling of the light curves of $\gamma$~Lup we used the orbital elements from Table~\ref{Tab-11} and the fundamental 
parameters $T_{\rm eff,1} ={}$20\,790~K and $\log g_1 ={}$3.94 from Appendix \ref{param-gl}. As in the case of $\nu$~Cen, the 
$T_{\rm eff,1}$ was fixed throughout but the final $\log g_1$ were taken from the W-D solutions after several iterations. The 
remaining details of the W-D modelling were also the same as in the case of $\nu$~Cen (see Section\,\ref{wd-ncen}), except that a 
third light was included in order to account for component B. The third light's brightness was kept as a free parameter because 
fixing it at a level consistent with the Hp magnitudes of the components (see Section\,\ref{targets}) resulted in a divergent 
solution. As in the case of $\nu$~Cen, the range of $M_1$ was assumed to span five times the uncertainty obtained in Appendix 
\ref{param-gl}. The results of the W-D modelling, listed in Table \ref{Tab-07}, bear many similarities to the results for 
$\nu$~Cen (Section\,\ref{wd-ncen}): (i) the mass ratio is low, $q = 0.15^{+0.09}_{-0.05}$, (ii) the W-D light-curves fit the 
normal points with very nearly the same overall standard deviation over the whole ($M_1$, $i$) grid, (iii) the position of the 
secondary relative to the ZAMS indicates its pre-MS status (see Fig.~\ref{FigHRpreMS}). The secondary's evolutionary status will 
be further discussed in Section\,\ref{discussion}.

\section{Discussion and conclusions}\label{discussion} 
\begin{figure*} 
\includegraphics[width=0.9\textwidth]{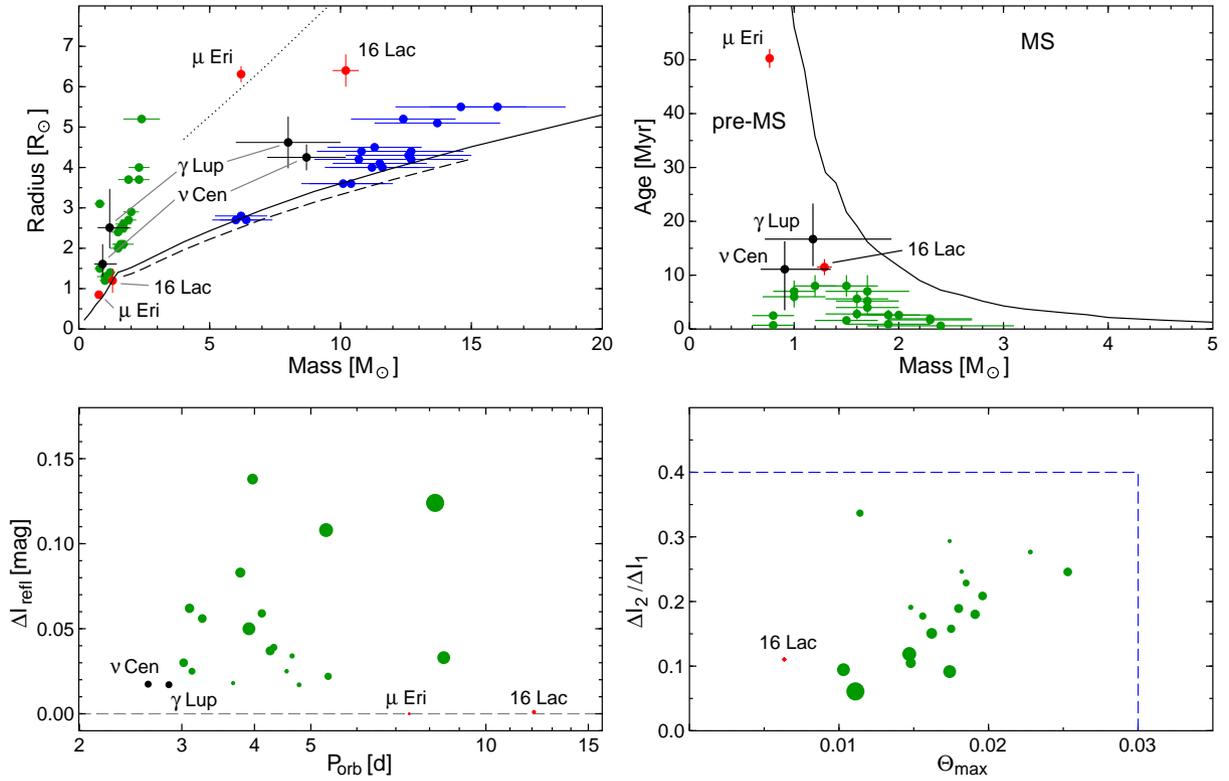} 
\caption{The stars discussed in the present paper in relation to the LMC NEBs. {\sl Upper left}: The mass-radius diagram for the 
primaries (blue) and the secondaries (green) of the LMC NEBs. The components of $\nu$~Cen and $\gamma$~Lup A are plotted as black 
dots, those of 16 (EN) Lac and $\mu$ Eri, as red dots. The solid line is the ZAMS mass-radius relation for the $Z=0.014$ models 
with no rotation from \citet{2012A&A...537A.146E} for $M\geq0.8$\,M$_\odot$, and from \citet*{2011A&A...533A.109T} for 
$M<0.8$\,M$_\odot$. The dashed line is the ZAMS relation for the non-rotating $Z=0.006$ models (i.e., those with the average 
metallicity of the LMC) from \citet{2013A&A...553A..24G}. Finally, the dotted line is the terminal-age main sequence (TAMS) 
relation for the above-mentioned \citet{2012A&A...537A.146E} models. {\sl Upper right}: The mass-age relation for the secondaries 
of the LMC NEBs, $\nu$~Cen, $\gamma$~Lup A, 16 (EN) Lac, and $\mu$ Eri. The line defines the border between the pre-MS and MS 
regions. The values of the pre-MS phase duration were derived from the models of \citet{2011A&A...533A.109T}. {\sl Lower left}: 
The range of the reflection effect in the $I$ band, $\Delta I_{\rm refl}$, versus orbital period, $P_{\rm orb}$ for the LMC NEBs 
(green), $\nu$~Cen and $\gamma$~Lup A (black), and 16 (EN) Lac (red). The diameters of the symbols are proportional to the stellar 
radii. {\sl Lower right}: The ratio of the eclipse depth versus the width of the wider eclipse, $\Theta_{\rm max}$, for the LMC 
NEBs and 16 (EN) Lac. For the latter, these parameters were derived from an eclipse light-curve obtained from the NASA Transiting 
Exoplanet Survey Satellite \citep[{\sc TESS},][]{2015JATIS...1a4003R} observations. The dashed lines contain the LMC NEBs region 
shown in the central panel of fig.~1 of \citet{MoDiS15}.}
\label{Fig-09cor3} 
\end{figure*}

Using stellar parameters and the results of the W-D modelling of the reflection effect in $\nu$~Cen and $\gamma$~Lup A, we 
concluded in Sections \ref{wd-ncen} and \ref{wd-glup} that the secondaries in both systems are in the pre-MS stage of evolution. 
Thus, as we already mentioned in Section \ref{intro}, the systems can be regarded as non-eclipsing counterparts of the NEBs, to be 
referred to in the following as NnonEBs. In order to strengthen this conclusion, we compare in Fig.\,\ref{Fig-09cor3} the radii, 
the age, and the range of reflection effect of $\nu$~Cen and $\gamma$~Lup A with those of the LMC NEBs using the parameters of 
$\nu$~Cen and $\gamma$~Lup A from Table \ref{Tab-07} (the ranges of the parameters are plotted as error bars) and those of the LMC 
NEBs from tables 1 and 2 of \cite{MoDiS15}. First of all, the radii of the secondaries are significantly larger than the zero-age 
main sequence (ZAMS) values for stars with the same masses (upper left-hand panel of the figure). A more convincing argument that 
$\nu$~Cen and $\gamma$~Lup A are indeed NnonEBs comes from the positions of their secondaries in the mass-age diagram (upper 
right-hand panel). The ages of the systems in the diagram were taken from Appendices \ref{param-nc} and \ref{param-gl}, where 
they are determined from the position of the primary components in the HR diagram in relation to the evolutionary tracks. As in 
the case of the LMC NEBs, both secondaries are located on the pre-MS side of the line which divides the pre-MS and MS regions. The 
two lower panels of Fig.\,\ref{Fig-09cor3} show parameters characterising the light curves. The NnonEBs $\nu$~Cen and $\gamma$~Lup 
A can be plotted only in the left-hand panel, in which they occupy a short-$P_{\rm orb}$ extension of the area occupied by the LMC 
NEBs. Also shown in Fig.\,\ref{Fig-09cor3} are two SB1 eclipsing binaries 16 (EN) Lac and $\mu$ Eri mentioned in 
Section\,\ref{intro}. 16 (EN) Lac is plotted with the parameters from \cite{MJ+15}, while $\mu$ Eri, with the parameters computed 
using the data from tables 3 and 4 and fig.~11 of \cite{MJ+13}. Because of the ratio of the radii $k = R_2/R_1$ of only $\approx 
0.25$ and a relatively long orbital period of 12.1\,d, the range of the reflection effect in 16 (EN) Lac is smaller than 5\,mmag. 
The pre-MS status of the star's secondary component is based on the analysis of the ground-based eclipse light-curve 
\citep{PJ88,MJ+15} and an unpublished eclipse light-curve obtained by one of the authors (MJ) from the NASA Transiting 
Exoplanet Survey Satellite \citep[{\sc TESS},][]{2015JATIS...1a4003R} observations. In the case of $\mu$ Eri, the orbital period 
is about 1.6 times shorter than that of 16 (EN) Lac but the reflection effect is below the detection threshold because of the 
small $k \approx 0.135$. Still, as can be seen from Fig.\,\ref{Fig-09cor3}, the pre-MS status of the secondary is evident. We 
conclude that 16 (EN) Lac and $\mu$ Eri should be regarded as bona fide NEBs. Note that since it was the reflection signature in 
the eclipsing light-curves which \citet{MoDiS15} used to single out the NEBs from the OGLE data, objects similar to 16 (EN) Lac 
and $\mu$ Eri would have passed undetected in their search.

As far as we are aware, $\nu$~Cen and $\gamma$~Lup A are the only known NnonEBs, i.e early B-type non-eclipsing SB systems 
such that (i) their observed orbital light-variation is caused solely by the reflection effect, and (ii) the secondary component 
is in a pre-MS evolutionary phase. In order to verify this and prepare the ground for a future photometric program, we 
have searched the literature for the light-variability information about all 169 B0\,--\,B5 non-eclipsing SB systems listed in The 
Ninth Catalogue of Spectroscopic Binary Orbits\footnote{http://sb9.astro.ulb.ac.be} \citep{Pourb+04}. Apart from $\nu$~Cen and 
$\gamma$~Lup A, we found only 17 systems with the orbital period equal to the light-variation period, all with primaries of 
spectral type B3 or earlier. In three cases the secondary component is known to be a compact object, a neutron star or a white 
dwarf, while in 13, the light curve is a double wave implying an ellipsoidal variation. Only one system, CX\,Dra, with an orbital 
period equal to 6.696\,d, exhibits a sinusoidal light-variation of this period, albeit with a large amount of scatter, and the 
phase relation between the RV and light variation characteristic of the reflection effect \citep{Koub+80}. The scatter is mainly 
due to the fact that CX~Dra is an interacting Be binary showing variations on time-scales from days to years. \citet{Koub+80} 
suggest that a combination of ellipsoidal and reflection effects is responsible for the orbital light-variation. However, using 
the well-known formula for the amplitude, $\delta m$, of the ellipsoidal light-variation \citep[see e.g.][]{Ru70} and the 
spectroscopic orbital parameters of the system \citep{Koub78} we find $\delta m \loa0.3$\,mmag. Thus, CX\,Dra meets condition (i). 
G.~D.~Penrod (1985, private communication to \citealt{Ho+92}) detected the lines of the secondary component and estimated the MK 
type to be F5\,III. The orbit of the secondary component was derived by \citet{Ho+92}. These authors concluded that the component 
is a mid-F luminosity III star filling its Roche lobe. This conclusion is at variance with the result of \citet*{Gu+84}, obtained 
by means of a thorough W-D modeling, that the system is detached. In any case, CX\,Dra does not meet condition (ii). We 
conclude that no other NnonEBs than $\nu$~Cen and $\gamma$~Lup A are known. However, there is a number of B0\,--\,B5 
non-eclipsing SB systems which in the Hipparcos Epoch Photometry have Hp ranges from two to several mmags but were not classified 
as periodic variables. Re-observed with satellite photometers, some of these systems may turn out to be NnonEBs.

As discussed in some detail by \cite{MoDiS15}, the study of nascent binaries may be of great importance for our understanding of 
the formation of low-mass-ratio binaries and the origin of such objects as type Ia supernovae, low-mass X-ray binaries and 
millisecond pulsars. The discussion would benefit from including Galactic counterparts of the LMC nascent binaries, eclipsing or 
otherwise. They should be searched for among young binaries. We have already found several candidates in young open clusters and 
associations but a detailed discussion of these objects is beyond the scope of this paper.

\section*{Acknowledgments}
We are indebted to Professor Jadwiga Daszy\'nska-Daszkiewicz for computing the evolutionary tracks used in Appendices 
\ref{param-nc} and \ref{param-gl}. This research has made use of the Washington Double Star Catalog maintained at the U.S.~Naval 
Observatory, the Aladin service, operated at CDS, Strasbourg, France, and the SAO/NASA Astrophysics Data System Abstract Service. 
APi, GM, and DM acknowledge support provided by the Polish National Science Center (NCN) grant No.~2016/21/B/ST9/01126. MR 
acknowledges support by the NCN grants No.~2015/16/S/ST9/00461 and 2017/27/B/ST9/02727. GH acknowledges support by the Polish NCN 
grant 2015/18/A/ST9/00578. APo was responsible for image processing and automation of photometric routines for the data registered 
by BRITE-nanosatellite constellation, and was supported by SUT grants: 02/140/SDU/10-22-01 and 02/140/RGJ21/0012. GAW acknowledges 
support in the form of a Discovery Grant from the Natural Science and Engineering Research Council (NSERC) of Canada. AJFM is 
grateful to NSERC (Canada) for financial aid. 

\section*{Data Availability}
The raw BRITE data are available from BRITE Public Data Archive (https://brite.camk.edu.pl/pub/index.html). The raw SMEI data are 
available via the link given in Section~\ref{sect:brite}. The processed (decorrelated) BRITE and SMEI data are available from the 
first author upon request. The other data are available from their respective public databases.

\bibliography{nuCen&gamLup}

\appendix
\section{Frequency analysis}\label{freqan}
\subsection{$\bnu$ Cen}\label{ap-nc}
\begin{figure} 
\centering
\includegraphics[width=0.9\columnwidth]{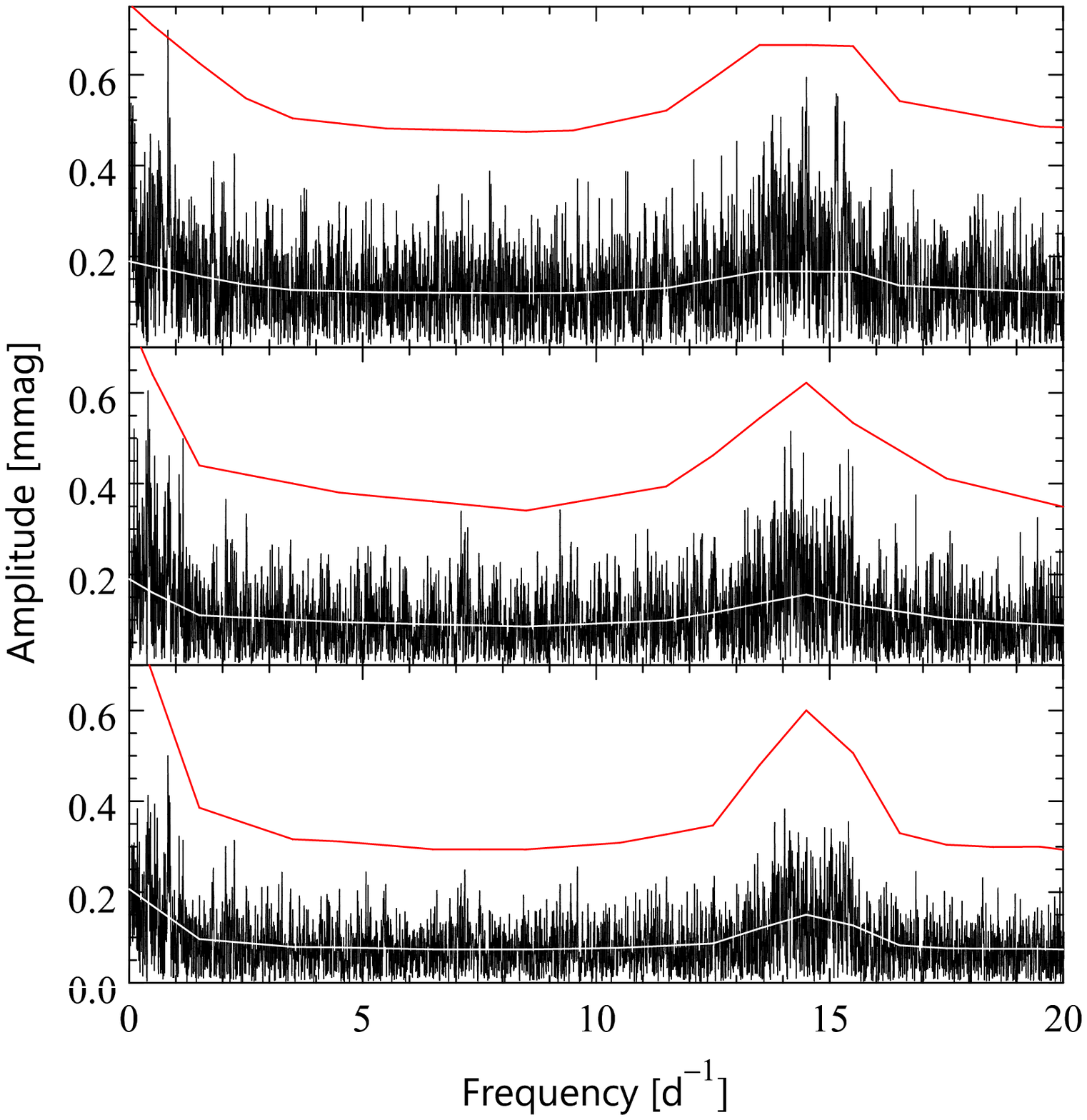} 
\caption{The periodograms of the blue (top), red (middle) and blue and red combined (bottom) BRITE magnitudes of $\nu$~Cen 
pre-whitened with the orbital frequency. The mean noise levels and four times those are indicated (the white and red lines, 
respectively).}
\label{Fig-A1} 
\end{figure}

For the purpose of frequency analysis, the reduced BAb and BLb magnitudes of $\nu$~Cen (Table \ref{Tab-01}) were combined into one 
data set of blue-filter magnitudes, and the reduced BTr and UBr magnitudes, into one data set of red-filter magnitudes. We shall 
refer to the two data sets as the red and blue data, respectively. The periodograms using these data were computed in the 
frequency range from 0 to 20\,d$^{-1}$. The highest peaks in the periodograms of both data sets occurred at the orbital frequency 
(Section\,\ref{porb-ncen}). The periodograms of the blue data, the red data and the blue and red data combined, pre-whitened with 
the orbital frequency, are shown in Fig.\,\ref{Fig-A1}. Also shown in the figure are the mean noise levels, $N$, and 4$N$, the 
popular detection threshold set by \citet{B+93}. As can be seen from the figure, there are no peaks in the periodograms exceeding 
0.7\,mmag. In particular, the blue and red amplitude at 2$f_{\rm orb}$ is equal to 0.1\,mmag, so that an ellipsoidal 
light-variation, if any, would have the amplitude less than or equal to 0.1\,mmag in both colours. At the frequencies 
corresponding to the putative $\beta$~Cephei-type periods of 0.1750, 0.1690156 and 0.1696401\,d, found by \citet{Raj77}, 
\citet{KS82} and \cite{ASR85}, the blue and red amplitudes do not exceed 0.2\,mmag. One peak in the top panel of 
Fig.\,\ref{Fig-A1} has $S/N$ slightly greater than 4, and a few ones in the middle panel have $S/N$ slightly smaller than 4, but 
none has a counterpart in the other panel. We conclude that in the data pre-whitened with the orbital frequency there are no 
periodic terms with an amplitude exceeding 0.7\,mmag. In the periodogram of the pre-whitened blue and red data combined (bottom 
panel), there are no peaks exceeding 0.5\,mmag. We thus confirm the results of \citet{Sh78}, \citet{PJM81} and \citet{SJ83} who 
found no short-period brightness variation, and strengthen that of \citet{Cuy+89} who found no variation, other than the orbital 
one, with an amplitude exceeding 2\,mmag (Section\,\ref{targets}). 

\subsection{$\bgamma$ Lup: the 2014 and 2015 data analyzed separately}\label{ap-gl1}
\begin{figure} 
\centering
\includegraphics[width=0.9\columnwidth]{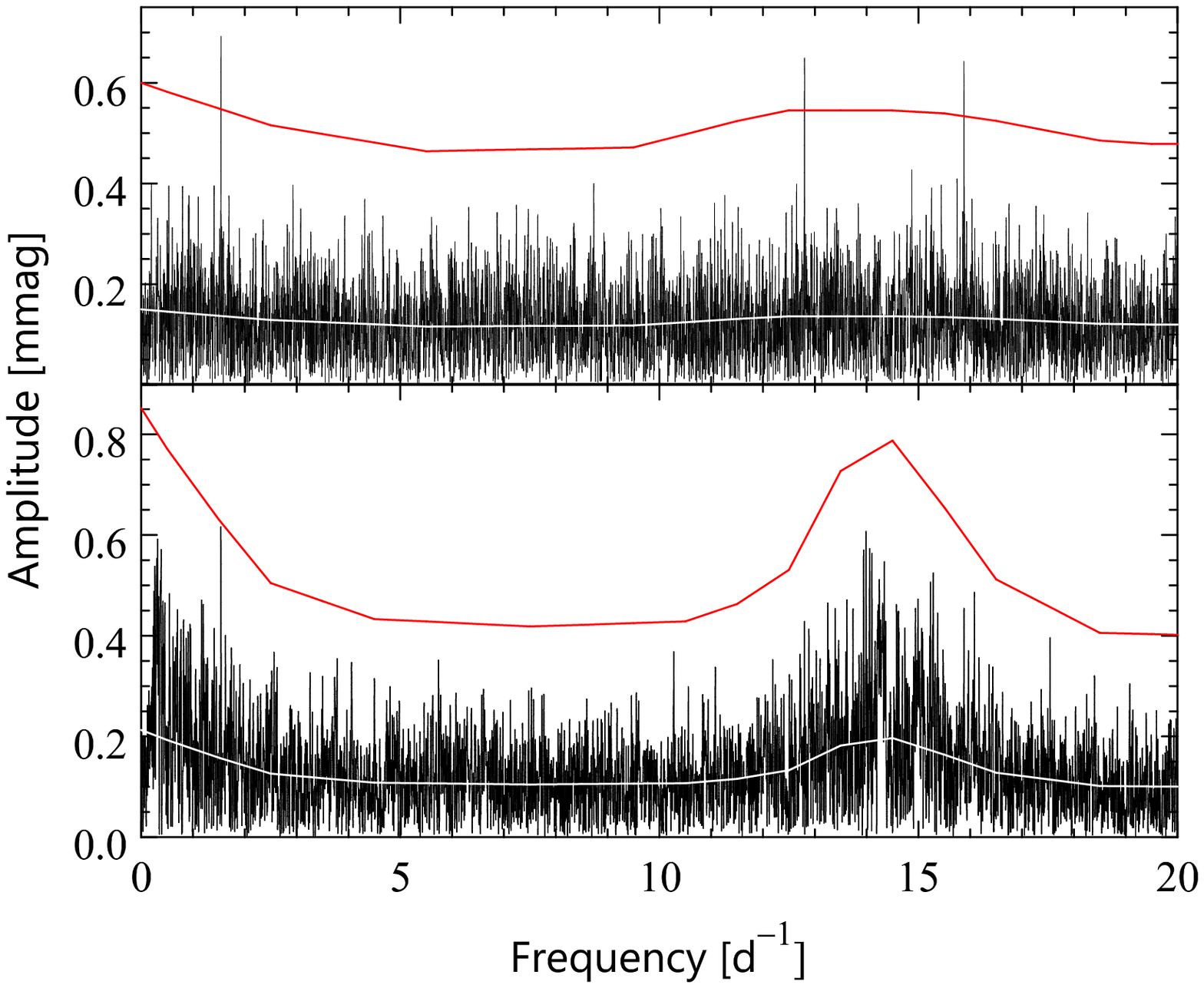} 
\caption{The periodograms of the 2014 blue (upper panel) and red (lower panel) BRITE magnitudes of $\gamma$~Lup pre-whitened with 
the orbital frequency. The mean noise levels and four times those are indicated (the white and red lines, respectively).}
\label{Fig-A2} 
\end{figure}

After combining the reduced 2014 BAb and BLb magnitudes of $\gamma$~Lup into one set of blue magnitudes, and the reduced 2014 UBr 
and BTr magnitudes, into one set of red magnitudes, we computed periodograms in the same way as we did previously for $\nu$~Cen 
(Section\,\ref{ap-nc}). In both periodograms, the frequency of the highest peak was equal to the orbital frequency of 
0.35090\,d$^{-1}$ to within 0.015 of the frequency resolution of the data. The periodograms of the blue and red magnitudes 
pre-whitened with the orbital frequency are shown in the upper and lower panels of Fig.\,\ref{Fig-A2}, respectively. In both 
panels, the highest peak occurs at 1.5389\,d$^{-1}$. At this frequency, the amplitude is equal to 0.7\,mmag and the 
signal-to-noise ratio $S/N >4$ in the upper panel, and the amplitude is equal to 0.6\,mmag and $S/N \approx 4$ in the lower panel. 
The blue to red amplitude ratio and the blue minus red phase difference of the 1.5389\,d$^{-1}$ sinusoid amount to 
1.13$\,\pm\,$0.16 and 0.43$\,\pm\,$0.14\,rad, respectively. These numbers are well within the range of values predicted for high 
radial-order, low harmonic-degree g-mode pulsations of B-stars models \citep[see e.g.][]{Tow02}. 

\begin{figure} 
\centering
\includegraphics[width=0.9\columnwidth]{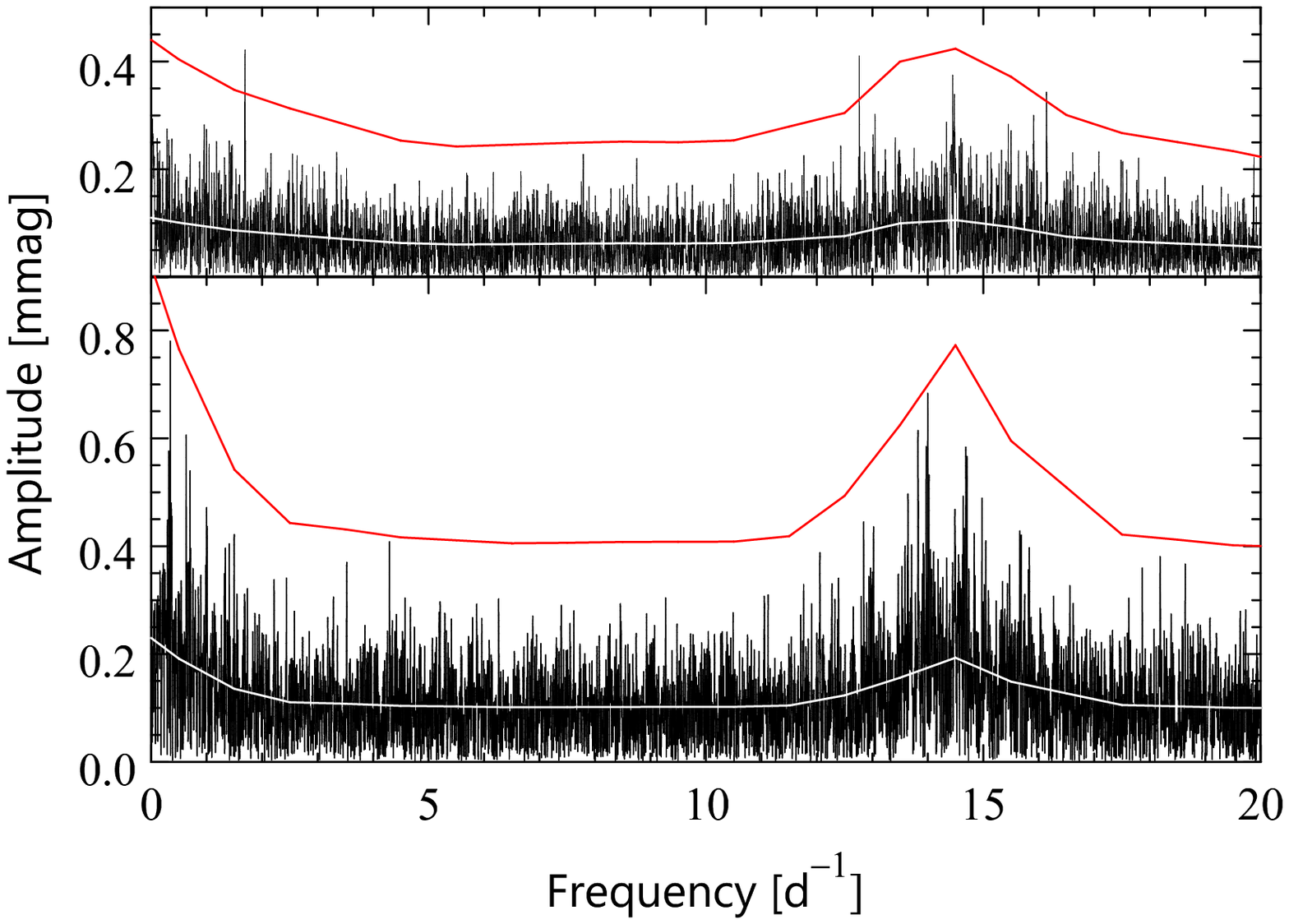} 
\caption{The same as in Fig.\,\ref{Fig-A2} but for the 2015 blue (upper panel) and red (lower panel) BRITE magnitudes of 
$\gamma$~Lup pre-whitened with the orbital frequency.}
\label{Fig-A3} 
\end{figure}

The periodograms of the 2015 blue and red magnitudes pre-whitened with the orbital frequency are shown in the upper and lower 
panel of Fig.\,\ref{Fig-A3}, respectively. The highest peak in the upper panel occurs at 1.6913\,d$^{-1}$, while the highest peak 
in the lower panel, at 0.3441\,d$^{-1}$. The former has $S/N >4$, while the latter, $S/N \approx 4$. However, none has a 
counterpart in the other panel, so that both are probably spurious. The 2014 1.5389\,d$^{-1}$ term was not present in 2015. 

\subsection{$\bgamma$ Lup: the 2014 and 2015 data combined}\label{ap-gl3}
\begin{figure} 
\centering
\includegraphics[width=0.9\columnwidth]{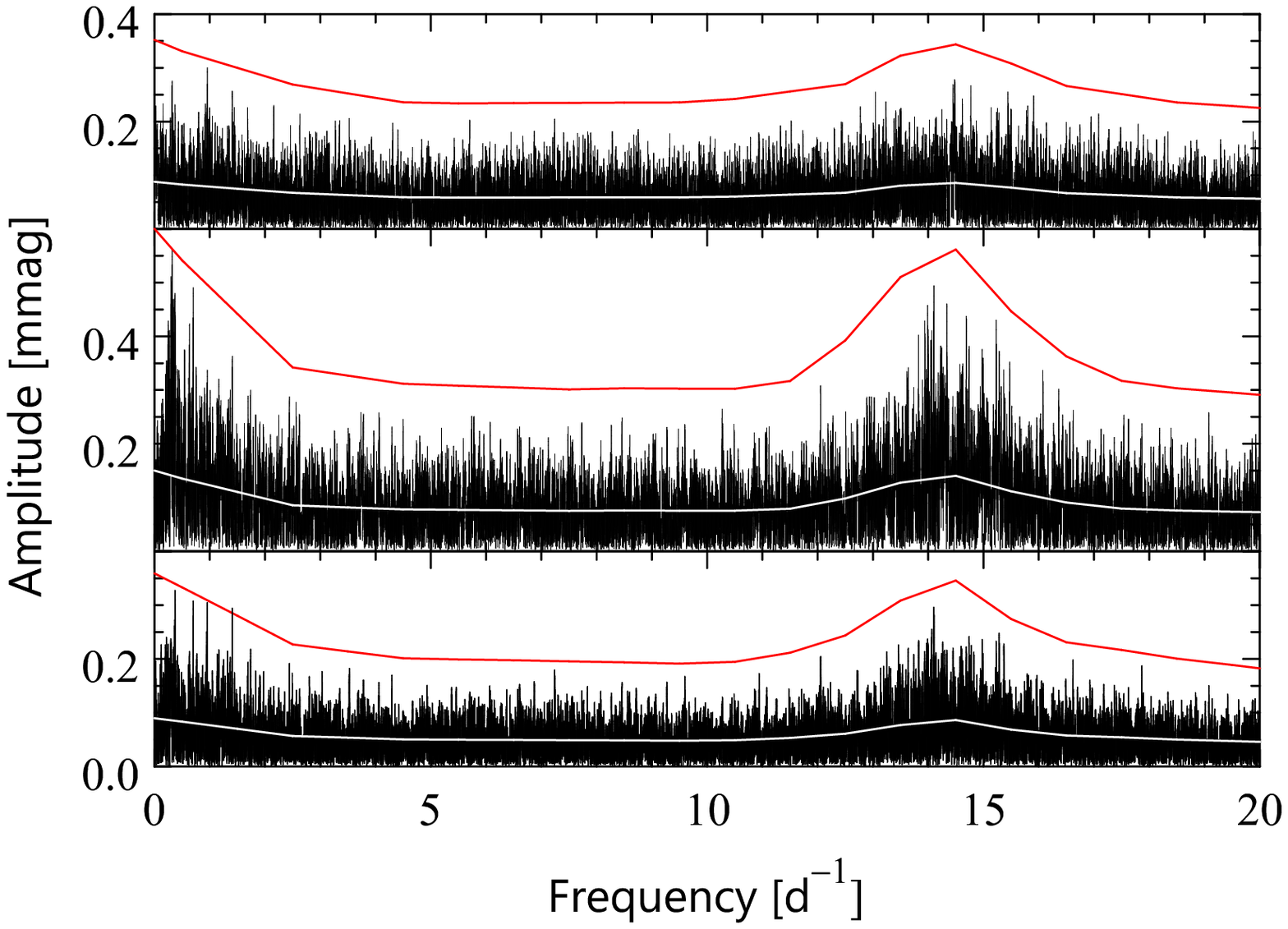}
\caption{The periodograms of the 2014 and 2015 blue (top), red (middle) and blue and red combined (bottom) BRITE magnitudes of 
$\gamma$~Lup, modified as explained in the text, pre-whitened with the orbital frequency. The mean noise levels and four times 
those are indicated (the white and red lines, respectively).}
\label{Fig-A4} 
\end{figure}

Before combining the 2014 and 2015 blue magnitudes into a single data set, we pre-whitened the 2014 blue magnitudes with the 
frequency of 1.5389\,d$^{-1}$, and the 2015 blue magnitudes with the frequency of 1.6913\,d$^{-1}$. Before combining the 2014 and 
2015 red magnitudes, we pre-whitened the 2014 red magnitudes with the frequency of 1.5389\,d$^{-1}$. This was done because 
combining the data without pre-whitening resulted in periodograms with peaks close to 1.5389 and 1.6913\,d$^{-1}$, although not so 
high as those in the 2014 and 2015 periodograms. The highest peaks in the periodograms of the combined 2014 and 2015 blue and red 
magnitudes occurred at the same frequency of 0.35088\,d$^{-1}$, a value very nearly equal to the orbital frequency. The 
periodograms of the blue data, the red data, and the blue and red data combined, pre-whitened with the orbital frequency, are 
shown in the top, middle and bottom panel of Fig.\,\ref{Fig-A4}, respectively. There are no peaks exceeding 4$N$ in the top and 
middle panels. In the top panel, the highest peak, with $S/N=3.6$, occurs at 0.9554\,d$^{-1}$. In the middle panel, the two 
highest peaks, with $S/N=3.9$ and 3.6, occur at 0.3176 and 0.7014\,d$^{-1}$. The latter frequency is close to 2$f_{\rm orb}$. The 
amplitude of the sinusoid of this frequency is equal to 0.5\,mmag and the phase differs from that of the $f_{\rm orb}$ sinusoid by 
an insignificant 0.2$\,\pm\,$0.2\,rad. Thus, the red orbital light-curve does not deviate within errors from a strictly sinusoidal 
shape. The same is true for the blue orbital light-curve because the amplitude of 2$f_{\rm orb}$ in the top panel of 
Fig.\,\ref{Fig-A4} amounts to 0.1\,mmag. Finally, the periodogram of the blue and red data combined (bottom panel of 
Fig.\,\ref{Fig-A4}) shows peaks close to the above-mentioned frequencies 0.3176, 0.7014 and 0.9554\,d$^{-1}$, and also peaks at 
0.3725 and 1.4077\,d$^{-1}$, all with $S/N \approx 4.0$. It is doubtful that any of these frequencies, except the 0.7014\,d$^{-1}$ 
one, represents a variation intrinsic to $\gamma$~Lup. Note that (i) because of the ambiguity mentioned in Section\,\ref{targets}, 
either component A or B would be responsible for any of the frequencies discussed above except $f_{\rm orb}$ and 2$f_{\rm orb}$ 
(Section 5.2), and (ii) the amplitude of a variation of component A would suffer from light dilution caused by component B, and 
vice versa. 

\section{Fundamental parameters}\label{param}
\subsection{$\bnu$ Cen}\label{param-nc}
We shall now derive the fundamental parameters of $\nu$~Cen. From the star's Str\"omgren indices $b-y$ and $c_1$ given by 
\citet{hm} we obtain $c_0=0.082$, $(b-y)_0=-0.109$, $E(b-y)=0.007$, and $E(B-V)=0.009$\,mag by means of the canonical method of 
\citet{c78}. From $c_0$ we get the effective temperature, $T_{\rm eff}=22\,267$\,K, and the bolometric correction, ${\rm 
BC}=-2.25$\,mag, using the calibration of \citet{ds77}, $T_{\rm eff}=22\,550$\,K using {\sc UVBYBETA}\footnote{A FORTRAN program 
based on the grid published by \citet{md85}. Written in 1985 by T.T.\ Moon of the University London and modified in 1992 and 1997 
by R.\ Napiwotzki of Universitaet Kiel \citep*[see][]{n93}.} and 22\,289~K using the calibration of \citet{SJ93}. The close 
agreement of these $T_{\rm eff}$ values is due to the fact that the three temperature calibrations rely heavily on the OAO-2 
absolute flux calibration of \citet{cod}. Taking a straight mean of the above three values we arrive at $T_{\rm eff} =22\,370$~K. 
Realistic standard deviations of the effective temperatures of early-type stars, estimated from the uncertainty of the absolute 
flux calibration, amount to about 3\,\% \citep{n93,j94} or 670~K for the $T_{\rm eff}$ in question. The standard deviation of BC 
we estimate to be 0.20\,mag.

The surface gravity of a B-type star can be obtained from its $\beta$ index. Taking the $\beta$ index of $\nu$~Cen from \citet{hm} 
and $c_0$ from the preceding paragraph, we get $\log g=3.76$ by means of {\sc UVBYBETA}. According to \citet{n93}, the uncertainty 
of the $\beta$-index surface gravities of hot stars is equal to 0.25\,dex; we shall adopt this value as the standard deviation of 
the star's $\log g$.

The revised {\em Hipparcos\/} parallax of $\nu$~Cen is equal to 7.47$\,\pm\,$0.17~mas \citep{vL}. Taking the star's $V$ magnitude 
from \citet{m91}, $E(B-V)$ from the first paragraph of this section and assuming $R_V=3.2$ we get $M_V=-2.25\,\pm\,0.05$\,mag. 
This value and BC yield $M_{\rm bol}=-4.50\,\pm\,0.21$\,mag and $\log L/{\rm L}_{\sun}=3.70\,\pm\,0.08$. In computing $\log 
L/$L$_{\sun}$, we assumed M$_{{\rm bol}{\sun}}=4.74$\,mag, a value consistent with BC$_{\sun}=-0.07$\,mag, the zero point of the 
bolometric-correction scale adopted by \citet{ds77}. In summary, the fundamental parameters of $\nu$~Cen are: $T_{\rm 
eff}=22\,370\,\pm\,670$\,K, $\log g=3.76\,\pm\,0.25$, $\log L/{\rm L}_{\sun}=3.70\,\pm\,0.08.$\footnote{Gaia's Early Data 
Release 3 (https://www.cosmos.esa.int/web/gaia /early-data-release-3) parallax of 8.05$\,\pm\,$0.35~mas yields $\log L/{\rm 
L}_{\sun}=3.63\,\pm\,0.09$.} Percentage-wise, $T_{\rm eff}$ is best constrained, while $\log g$, the worst. In the {\sc PASTEL} 
catalogue of stellar atmospheric parameters \citep{Soub16} one finds $T_{\rm eff}=22\,570\,\pm\,1810$\,K, a value obtained by 
\citet{So95} from the continuum between 320 and 360\,nm. No $\log g$ of $\nu$~Cen is listed in the catalogue.

\begin{figure} 
\includegraphics[width=\columnwidth]{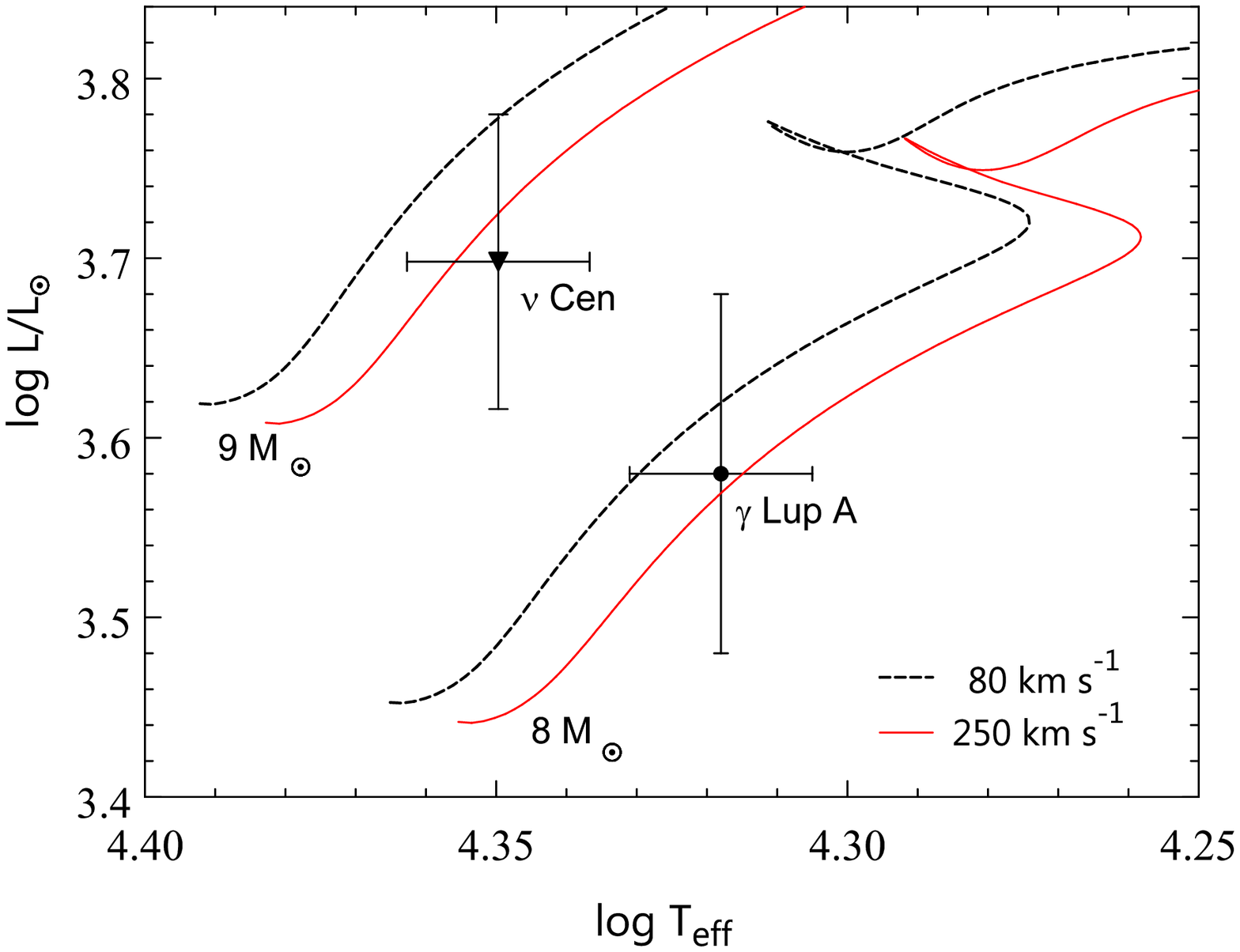} 
\caption{The primary components of $\nu$~Cen and $\gamma$~Lup A (inverted triangle and filled circle with error bars, 
respectively) in the HR diagram plotted using the effective temperature derived from photometric indices, and the luminosity from 
the $V$ magnitude, photometric indices and the {\em Hipparcos\/} parallax (see Section\,\ref{param-nc} for $\nu$~Cen and 
\ref{param-gl} for $\gamma$~Lup A). The lines are the $X=0.7$, $Z=0.015$ evolutionary tracks for 8 and 9\,M$_{\sun}$ and the 
rotation velocity listed in the bottom right corner.}
\label{Fig-B1cor} 
\end{figure}

Since the bolometric magnitude of the secondary component of $\nu$~Cen is at least 7.4\,mag fainter than that of the primary 
(Table~\ref{Tab-07}), so that it is at least 5\,mag fainter in $V$ than the primary, the fundamental parameters $T_{\rm eff}$, 
$\log L/$L$_{\sun}$ and $\log g$ just derived pertain to the latter; in Section\,\ref{wd-ncen} we refer to them as $T_{\rm 
eff,1}$, $\log L_1/$L$_{\sun}$ and $\log g_1$. Using the first two parameters, we plot the primary component in 
Fig.\,\ref{Fig-B1cor}. Also shown in the figure are evolutionary tracks computed by means of the Warsaw-New Jersey evolutionary 
code \citep[see e.g.][]{P+98}, assuming no convective-core overshooting, the initial abundance of hydrogen $X=0.7$, the 
metallicity $Z= 0.015$, the OPAL equation of state \citep{RN02} and the OP opacities \citep{S05} for the heavy element mixture of 
\citet{A+09}. The tracks were kindly provided by Professor J.\ Daszy\'nska-Daszkiewicz. As can be seen from Fig.\,\ref{Fig-B1cor}, 
the primary component of $\nu$~Cen is in the MS stage of evolution. The evolutionary mass and age, estimated from the position of 
the inverted triangle in the figure relative to the $V_{\rm rot}=80$\,km\,s$^{-1}$ evolutionary tracks, are equal to 
8.7$\,\pm\,$0.3\,M$_{\sun}$ and 11.1$^{+5.2}_{-7.6}$~Myr. The most recent value of $V_{\rm rot} \sin i$, measured on high 
signal-to-noise-ratio spectrograms, is equal to 65$\,\pm\,$6\,km\,s$^{-1}$ \citep{BV97}, so that the $V_{\rm rot}$ value we 
assumed corresponds to the inclination of the rotation axis of 54$^{+8}_{-6}$\,deg. We believe that in view of the large errors of 
$\log T_{\rm eff,1}$ and $\log L_1/$L$_{\sun}$, the contribution of the uncertainty in $V_{\rm rot}$ to the final error of the 
evolutionary mass and age is negligible. The evolutionary age we derived is smaller than the MS turnoff age of 17$\,\pm\,$1~Myr 
obtained for UCL by \citet*{MML02}, but the difference is still within the errors. Our value is close to the age of $\sim$10~Myr 
obtained from the strength of Li 670.8\,nm absorption line by \citet*{SZB12}. 

\subsection{$\bgamma$ Lup}\label{param-gl}
We shall now derive fundamental parameters of $\gamma$~Lup A assuming that the components A and B, because of their very nearly 
equal brightness (Section\,\ref{targets}), have identical photometric indices, equal to the observed, combined values. From the 
star's Str\"omgren indices $b-y$ and $c_1$ \citep{hm} we get $c_0=0.141$, $(b-y)_0=-0.103$, $E(b-y)=0.006$ and $E(B-V)=0.008$\,mag 
by means of the canonical method of \citet{c78}. Then, using the same procedures and calibrations as in the case of $\nu$~Cen 
(Section\,\ref{param-nc}), we obtain $T_{\rm eff}=20\,790\,\pm\,620$\,K, ${\rm BC}=-2.10\,\pm\,0.20$\,mag and $\log 
g=3.94\,\pm\,0.25$. In order to derive component's A logarithmic luminosity, $\log (L_{\rm A}/{\rm L}_{\sun})$, we first computed 
the $V$ magnitudes from the Hp magnitudes (Section\,\ref{targets}) using the \citet{PH98} transformation with the $B-V$ and $U-B$ 
indices from \citet{m91}. The result is $V_{\rm A}=3.468$ and $V_{\rm B}=3.582$\,mag. The combined $V$ magnitude of $\gamma$~Lup 
computed from $V_{\rm A}$ and $V_{\rm B}$ is 2.771\,mag, in good agreement with $V=2.780\,\pm\,0.013$\,mag given by \citet{m91}. 
Moreover, the B {\em minus} A $V$-magnitude difference implies an insignificant $B-V$ difference of 0.005\,mag, consistent with 
the assumption made at the beginning of this paragraph. From the revised {\em Hipparcos\/} parallax, equal to 
$7.75\,\pm\,0.50$\,mas \citep{vL}, the $E(B-V)$ derived above and $R_V=3.2$ we get $M_V^A=-2.11\,\pm\,0.14$\,mag. This value and 
BC yield $M_{\rm bol}^A=-4.21\,\pm\,0.24$\,mag. In summary, the fundamental parameters of $\gamma$~Lup A are: $T_{\rm 
eff}=20\,790\,\pm\,620$\,K, $\log g=3.94\,\pm\,0.25$ and $\log (L/{\rm L}_{\sun})=3.58\,\pm\,0.10$. As in the case of $\nu$~Cen, 
$T_{\rm eff}$ is best constrained, while $\log g$, the worst.

The secondary component of $\gamma$~Lup A is about 4.5 to 5.5\,mag fainter than the primary (Table~\ref{Tab-07}). Thus, the 
fundamental parameters $T_{\rm eff}$, $\log L/$L$_{\sun}$ and $\log g$ just derived pertain to the latter; in 
Section\,\ref{wd-glup} we refer to them as $T_{\rm eff,1}$, $\log L_1/$L$_{\sun}$ and $\log g_1$. Using the first two parameters, 
we plot the primary component in Fig.\,\ref{Fig-B1cor}. As can be seen from the figure, the primary component falls very nearly on 
the MS branch of the 8\,M$_{\sun}$, $V_{\rm rot}=250$\,km\,s$^{-1}$ evolutionary track. Its evolutionary mass and age, estimated 
from the position of the circle relative to the track, are equal to 8.0\,$\pm$\,0.4~M$_{\sun}$ and $16.7^{+5.0}_{-6.6}$~Myr. 
According to \citet{GG05}, $V_{\rm rot} \sin i$ of $\gamma$~Lup is equal to $236\,\pm\,5$\,km\,s$^{-1}$, so that the $V_{\rm rot}$ 
value we assumed is close to a lower limit because it corresponds to the inclination of the rotation axis of $71^{+4}_{-3}$\,deg. 
However, the evolutionary mass and age are rather insensitive to $V_{\rm rot}$ (see below). Unlike the case of $\nu$~Cen, the 
evolutionary age we derived is very nearly equal to the UCL's MS turnoff age of $17\,\pm\,1$\,Myr \citep{MML02}. Component B is 
not plotted in Fig.\,\ref{Fig-B1cor} because it would almost coincide with A. Its evolutionary mass and age would be very nearly 
equal to those of A, provided it had $V_{\rm rot}=250$\,km\,s$^{-1}$. If $V_{\rm rot}=0$\,km\,s$^{-1}$, the evolutionary mass and 
age would become 7.8\,$\pm$\,0.4~M$_{\sun}$ and $17.5^{+5.0}_{-6.6}$\,Myr. Since the difference between the evolutionary masses is 
much smaller than 1$\sigma$, we conclude that the sum of evolutionary masses of the spectroscopic primary of A and that of B 
is equal to about 15.9\,$\pm$\,0.6~M$_{\sun}$. Assuming the mass of the spectroscopic secondary of A to be equal to 
1.3$\,\pm\,$0.3~M$_{\sun}$, i.e. the middle of the range of $M_2$ in Table~\ref{Tab-07} with an estimated uncertainty, and 
assuming that B is single, we get 17.2$\,\pm\,$0.7~M$_{\sun}$ for the overall mass of the AB system, and 0.86$\,\pm\,$0.12 for the 
B to A mass ratio. The overall mass of the AB system is thus about 5.5\,M$_{\sun}$ smaller than $M_{\rm A} + M_{\rm B}$ derived in 
Section\,\ref{casb} for $P_{\rm AB}=188.8$\,yr. Hopefully, this discrepancy will be removed when more measurements of the position 
angle and angular separation of $\gamma$~Lup B relative to A, and therefore a better visual orbit of $\gamma$~Lup AB than that 
given in Table~\ref{Tab-08} become available. 

\bsp

\label{lastpage}
\end{document}